\pdfoutput=1

\documentclass[preprint,12pt]{elsarticle}

\usepackage{amssymb}
\usepackage{natbib}
\usepackage[font=scriptsize]{caption}
\usepackage{graphicx, subfig}
\usepackage[colorlinks]{hyperref}
\usepackage{float}
\usepackage{color}
\usepackage{threeparttable}

\journal{Journal of Informetrics}

\begin{document}

\begin{frontmatter}



\title{Become a better you: correlation between the change of research direction and the change of scientific performance}

\author[label1]{Xiaoyao Yu}
\author[label2,label3]{Boleslaw K. Szymanski}
\author[label1]{Tao Jia\corref{mycorrespondingauthor}}
\cortext[mycorrespondingauthor]{Corresponding author}
\ead{tjia@swu.edu.cn}
\address[label1]{College of Computer and Information Science, Southwest University, Chongqing, 400715, P. R. China}
\address[label2]{Network Science and Technology Center, Rensselaer Polytechnic Institute, Troy, NY, USA}
\address[label3]{Społeczna Akademia Nauk, Łódź, Poland}

\begin{abstract}
It is important to explore how scientists decide their research agenda and the corresponding consequences, as their decisions collectively shape contemporary science. There are studies focusing on the overall performance of individuals with different problem choosing strategies. Here we ask a slightly different but relatively unexplored question: how is a scientist's change of research agenda associated with her change of scientific performance. Using publication records of over 14,000 authors in physics, we quantitatively measure the extent of research direction change and the performance change of individuals. We identify a strong positive correlation between the direction change and impact change. Scientists with a larger direction change not only are more likely to produce works with increased scientific impact compared to their past ones, but also have a higher growth rate of scientific impact. On the other hand, the direction change is not associated with productivity change. Those who stay in familiar topics do not publish faster than those who venture out and establish themselves in a new field. The gauge of research direction in this work is uncorrelated with the diversity of research agenda and the switching probability among topics, capturing the evolution of individual careers from a new point of view. Though the finding is inevitably affected by the survival bias, it sheds light on a range of problems in the career development of individual scientists.
\end{abstract}


\end{frontmatter}

\section{Introduction}
The availability of large scale data set and computational tools makes it easier than ever to quantitatively probe the science at different scales, from papers \citep{wang2013quantifying, hu2020describing, chen2021exploring, mukherjee2017nearly} to individual scientists \citep{way2017misleading, liu2018hot, bu2018understanding, zhang2017identifying}, and from research teams \citep{wu2019large, ma2020mentorship, alshebli2018preeminence} to institutions and nations \citep{king2004scientific, liu2020dominance, zhao2020investigation, zuo2018more, chen2020rank, huang2020comparison}. As the development of science is driven by scientists' involvements in different research topics, it is crucial to understand how they decide their research directions and what are the consequences of their collective decisions. Such choices affect not only individual careers but also collectively shapes contemporary science. The conflict in topic choosing is vividly depicted as ``The essential tension'' \citep{kuhn1977the} that scientists often need to decide whether to explore a new field or to exploit familiar topics \citep{foster2015tradition}. 

By carefully selecting the control group or by comparing a group of elites with the average, quantitative studies have provided us with a consistent conclusion. Those who focus on a narrow research agenda tend to secure a steady scientific output and gain more overall citations \citep{amjad2018measuring, zeng2019increasing, pramanik2019migration} whereas those who take the risk to change topics are likely to produce ``hit'' papers or highly innovative outcomes \citep{leahey2017prominent, foster2015tradition, azoulay2011incentives}. However, existing works usually focus on a group of selective individuals whose scientific achievements are beyond a certain threshold. It is still unclear what would happen to a typical scientist who crosses the boundary and enters a new field. Moreover, the comparison is usually based on the overall performance and made between two distinct groups of individuals. The change of performance, measured by the past and current performance of one's own, is rarely investigated.

Here, we ask a set of simple but relatively unexplored questions: If one changes the research agenda and succeeds in the new field, is she going to publish better research works than she used to do? Is she going to publish faster or slower compared with her past publication speed? Furthermore, if she were to venture further, would her benefits/disadvantages increase or decline? These questions focusing on the change of the scientific performance of an individual are not fully addressed in existing works, to the best of our knowledge. We understand that individual scientists are inherently different from each other, affected by confounding factors such as scientific training \citep{clauset2015systematic}, gender \citep{mauleon2006productivity, huang2020historical}, age \citep{jones2011age, petersen2011quantitative}, mobility \citep{robinson2019many, petersen2018multiscale, zhao2020investigation}, prestige of institution in which they work \citep{way2019productivity}, and more. While some factors can be controlled, many of them are hard to control. Therefore, we choose to use a large size of samples which may allow us to balance out other factors and reach preliminary answers to these questions. We use data about the publication in the American Physical Society (APS) journals and apply the approach by \cite{jia2017quantifying,aleta2019explore} to quantitatively measure the extent of research direction change. The method does not just provide a binary classification that tells if a scientist changed the research direction or not. Instead, it gauges the distance between two vectors characterizing the topics of a paper set, giving rise to a continuous measure of the direction change. Using the publication records of over 14,000 scientists, we find that the research direction change is positively correlated with the increase of impact: those with a bigger change in the research direction demonstrate not only a higher probability to increase citations of the publications, but also the relative magnitude of the citation gain. In contrast, the relationship between the research direction change and productivity change is neutral. Scientists who stay on the same topic do not produce faster than their colleagues venturing into new areas. We observe similar patterns when varying the criteria of data filtering, which supports the robustness of our conclusion. 

We also carefully compare our results with some recent findings. We provide evidence that the metric used in our study is totally uncorrelated with the one in \cite{zeng2019increasing} and the quality associated with the direction change is also different. Hence, the two studies quantify the effects of direction change from two distinct perspectives. We compare the direction change with the diversity change of an individual's research agenda \citep{wang2015interdisciplinarity, leahey2017prominent} and find that these two quantities are uncorrelated. Therefore, the gain of citation is not from the benefit of conducting diverse or interdisciplinary research. We also control the field citation and find that our conclusion still holds. In other words, the positive correlation between direction change and impact change is not from the move to a hot field. Taken together, we show the correlation between the research direction change and the performance change among a large number of physicists, which advances our understanding of the career development of individual scientists.

\section{Materials and Methods}

\subsection{Data set}
The APS data is publicly available at http://journals.aps.org/datasets, which consists of around 300,000 scientific papers authored by around 200,000 scientists together with their citation records covering years from 1976 to 2009. The information includes a paper's author list, its publication time, its reference list, and more. To classify papers belonging to each author, we perform the name disambiguation following procedures in previous works \cite{deville2014career, zeng2019increasing, aleta2019explore}. The name disambiguated data are roughly the same as one we used in another work \cite{jia2017quantifying}. Still, some author names are so general that individual authors can hardly be distinguished. For this reason, we calculate the Shannon entropy of topics in an author's publications and filter out those outliers who participate in unexpected highly diverse topics. We first find the topic vector $g$ based on the all papers under one author (see below for composing topic vector) and then calculate Shannon entropy as $H =\sum^{67}_{j=1} - x_{j}log(x_{j})$ where $x_{j}$ is the $j^{th}$ element of the vector $g$. Following \cite{jia2017quantifying}, we set $H=2.5$ as the cutoff value.

\subsection{Composing topic vector}
We make use of the Physics and Astronomy Classification Scheme (PACS) code firstly proposed by the American Institute of Physics in 1975. The PACS code contains 6 digits in a format like ``ab.cd.ef'' that points to a specific and specialized area in physics. The order of these digits is related to the hierarchical structure of the topics. For example, the code ``82.39.Rt'' corresponds to ``Reactions in complex biological systems'' which is a subtopic under the first level topic ``82'': physical chemistry and chemical physics. Totally, there are 67 topics given by the first two digits of the PACS code, representing 67 sub-disciplines in modern physics. Over 90\% of papers published after 1985 are labeled by one to three PACS codes.

For each paper, we count the occurrence of the first level topic given by the first two digits of a PACS code. The number of occurrences is further normalized by the total number of PACS codes in a paper. Then for this paper, we can build a vector $A = (a_{1}, a_{2}, ...,a_{j},... a_{67})$ where the element $a_j$ is the fractionalized occurrence of each topic in this paper. For a set of $m$ papers, we can average the $m$ vectors of each paper and reach a topic vector $g$ representing the weight coverage of 67 topics that the set of $m$ papers demonstrates. If the $m$ papers are authored by one scientist, the topic vector $g$ then provides a good proxy of her research direction, capturing not only the collection of topics but also the level of involvements she studies in the $m$ papers. An example of composing a topic vector is illustrated in Fig. \ref{fig1}a.

\subsection{Measuring the research direction change}
We sequence a scientist's $n$ papers according to their publication time. We then select two sets of papers, each with size $m$, in a scientist's early and late career, from which we can build two topic vectors. Denote $g_i$ by the topic vector of the early publications and $g_f$ by the late publications. Comparing the difference between these two vectors allows us to quantify the degree of research direction change $J_o$. Using cosine similarity, we have
\begin{equation}
	J_{o} = 1 - \frac{g_i\cdot g_f }{\left|\left|g_i\right|\right| \left|\left|g_f\right|\right|}\label{(1)}.
\end{equation}
$J_{o}$ varies between 0 and 1, where $J_{o} = 0$ means the author studies the same topic in the early and late career with identical involvements, $J_{o} = 1$ corresponds to the largest change as the topics in the two sets of papers have no overlap.

\begin{figure}[H]
	\begin{center}
		\includegraphics[width=13.5cm]{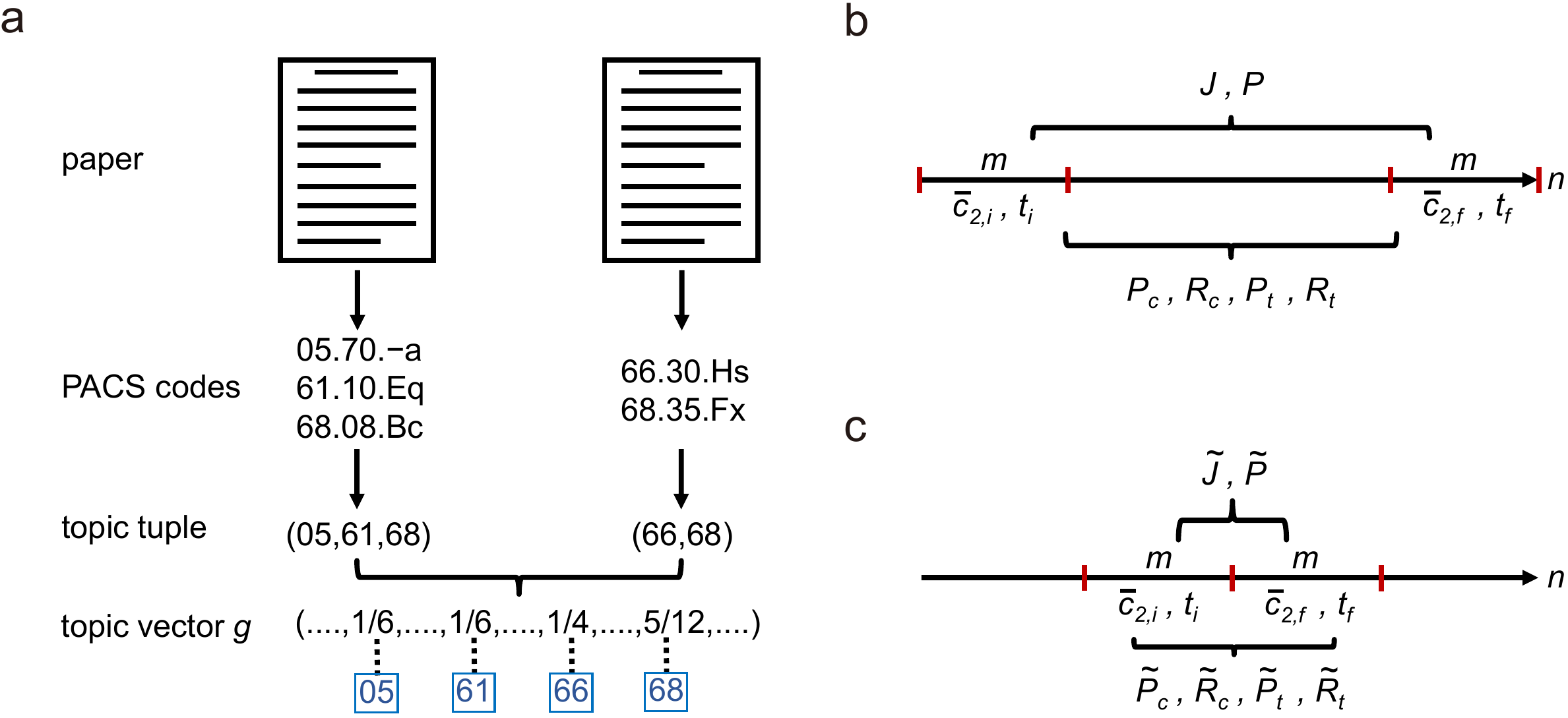}
		\caption{
			({\bf a}) An example demonstrating the procedure to compose topic tuple and topic vector \textit{g}. For two topic tuples (66, 68) and (05, 61, 68), the element value in $g$ of topic 66 is calculated as $\frac{1/2+0}{2} = \frac{1}{4}$, as it appears once in one topic tuple and is not included in the other. The element value in $g$ of topic 68 is calculated as $\frac{1/2+1/3}{2} = \frac{5}{12}$, as it appears once in each of the topic tuples. Similarly, the element values in $g$ of topic 05 and 61 are calculated as $\frac{0+1/3}{2} = \frac{1}{6}$.
			({\bf b}) The scenario that takes the first and the last $m$ papers in a scientist's publication sequence to obtain the direction change $J$, its distribution $P$, the growth fraction and growth rate of the scientific impact and productivity $P_c$, $R_c$, $P_t$, and $R_t$.
			({\bf c}) The scenario that uses two adjacent sequences of $m$ papers randomly chosen from a scientist's publication sequence. Correspondingly, the quantities obtained are denoted by $\tilde{J}$, $\tilde{P}$, $\tilde{P}_c$, $\tilde{R}_c$, $\tilde{P}_t$, and $\tilde{R}_t$.
			\label{fig1}}
	\end{center}
\end{figure}

To make sure that the results are not affected by how the two sets of papers are selected, we consider two distinct scenarios. First, we select two series of $m$ papers maximally separated in a scientist's publication sequence, \textit{i.e.} the first and last $m$ papers, capturing the change across the whole recorded career (Fig. \ref{fig1}b). The corresponding direction change is denoted by $J$. As one publication sequence defines one unique career, we only have one measure of $J$ for each author. 

Additionally, we also consider two consecutive sets of papers beginning at a random choosing paper to eliminate potential effects caused by the gap between the two sets of publications (Fig. \ref{fig1}c). The direction change measured is denoted by $\tilde{J}$. As an author may have published more than $2m$ papers, there are multiple measures of $\tilde{J}$ for each author. Therefore, for all measures associated with this scenario, we randomly pick a beginning paper for each author, get the measures needed (such as $\tilde{J}$, $\tilde{P}_t$ and so on), and calculate the statistics across the population. Then we repeat this procedure 2000 times. We report the mean value and the error bar that corresponds to the standard deviation. Though the value and the exact distribution of $J$ and $\tilde{J}$ are different, it is also noteworthy that the two quantities are highly correlated. The Pearson correlation between $J$ and the average value $\langle \tilde{J} \rangle$ of an individual scientist is over 0.99 (Fig. \hyperlink{figS1}{S1}). It implies that the big research direction change is likely carried out incrementally throughout one's career.

Due to the way we quantify the research direction change, only authors with no fewer than $2m$ papers can be included in our analysis. In the main text, we report results based on $m = 8$, corresponding to 14,726 scientists who authored at least 16 papers. We repeat the same analysis with different \textit{m} values ($m = 7$ and $m = 9$, Section \hyperlink{Note.S2}{S2}). The same pattern is observed, suggesting that our findings do not depend on the choice of $m$.

\section{Results}

\subsection{Correlation between the direction change and impact change}
We apply two distinct approaches to quantify the impact change. First, we group scientists whose direction changes fall into a small range ($J \in (0.025, 0.075]$ for instance) and measure the percentage of scientists within this group whose research impact have increased. In particular, we count the number of citations a paper receives within two years after its publication and normalize this value by the average citations of papers published in the same year, giving rise to a normalized citation measure $c_{2}$ that takes into account the citation inflation \citep{radicchi2008universality,sinatra2016quantifying,petersen2019methods,huang2020historical,huang2020patent} (Section \hyperlink{Note.S1}{S1}). We then measure the average value of $c_{2}$ for the $m$ papers in an author's early and late publication list which is used to quantify research direction change. Denote $\bar{c}_{2,i}$ and $\bar{c}_{2,f}$ by the average citation of the two paper sets, respectively (Figs. \ref{fig1}b-c). If an author's late publications on average receive more citations than her early ones ($\bar{c}_{2,f} > \bar{c}_{2,i}$), we consider this author's research impact has increased. The percentage ${P}_{c}$ is then calculated as the ratio between the number of authors whose $\bar{c}_{2,f} > \bar{c}_{2,i}$ and the number of authors whose direction change fall within the range ($J-0.025,J+0.025$]. For the direction change $\tilde{J}$, we obtain $\tilde{P}_c$ with a similar approach.

\begin{figure}[H]
	\begin{center}
		\includegraphics[width=13.5cm]{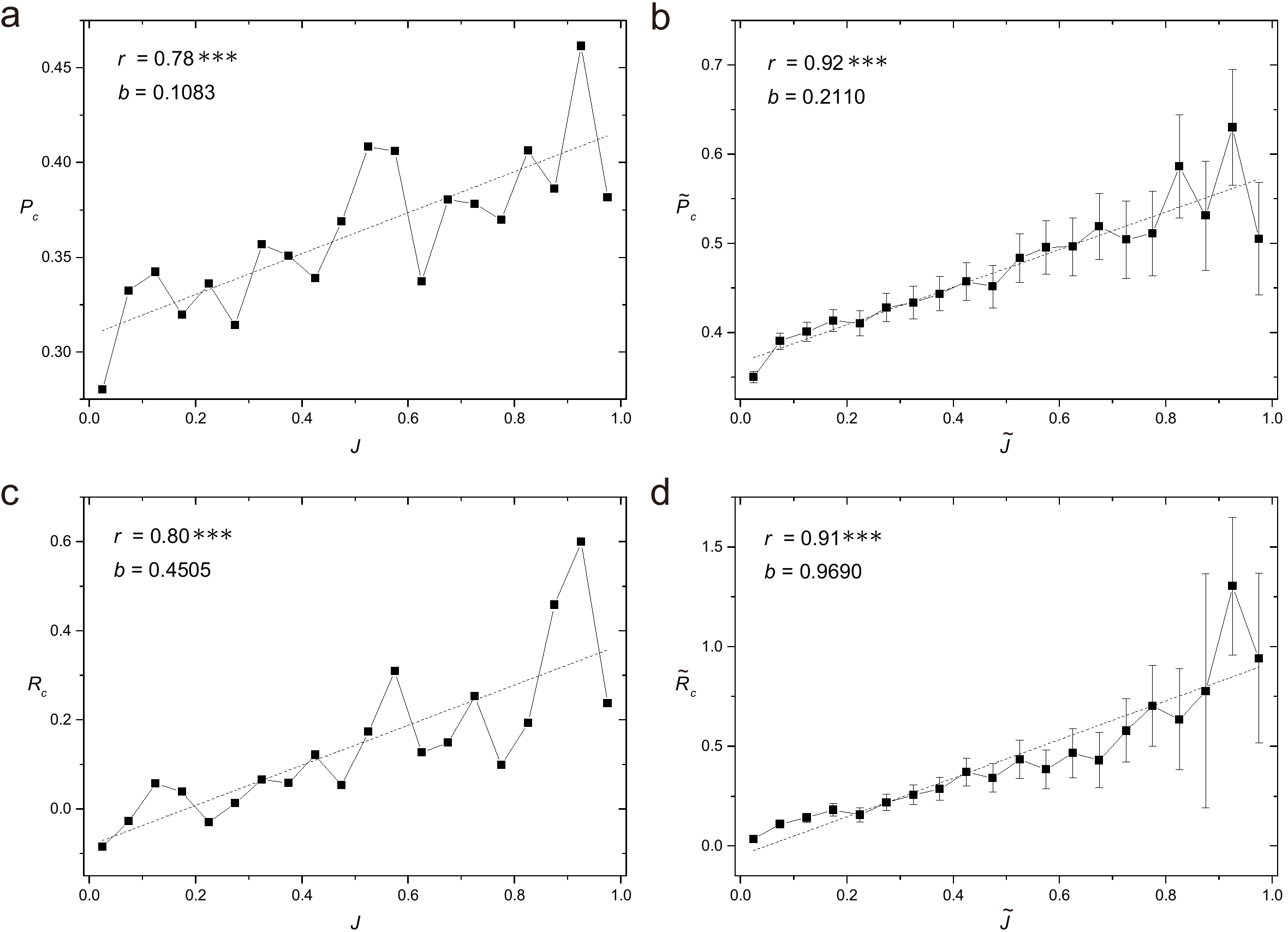}
		\caption{
			({\bf a}) $P_c$ conditioning on the range ($J-0.025,J+0.025$] is positively correlated with $J$. The dashed line represents the linear regression.
			({\bf b}) The average $\tilde{P}_c$ conditioning on the range ($\tilde{J} - 0.025, \tilde{J} + 0.025$] is positively correlated with $\tilde{J}$.
			({\bf c}) The average $R_c$ conditioning on the range ($J-0.025,J+0.025$] is positively correlated with $J$.
			({\bf d}) The average $\tilde{R}_c$ conditioning on the range ($\tilde{J} - 0.025, \tilde{J} + 0.025$] is positively correlated with $\tilde{J}$. At the boundary $J=0$ and $\tilde{J}=0$, the range [0, 0.05] is used, and the same boundary condition applies in all the analyses of $J$ and $\tilde{J}$. The scatter plots of $P_c$, $\tilde{P}_c$, $R_c$, and $\tilde{R}_c$ are displayed in Fig. \protect\hyperlink{figS2}{S2}. The value of $b$ is defined as the slope of the corresponding linear regression function (The dashed line). *** $p < 0.001$, ** $p < 0.05$, * $p < 0.1$ ($t$-test for Pearson coefficient $r$). Error bars represent the one standard deviation of the mean.
			\label{fig2}}
	\end{center}
\end{figure}

The interpretation of these two measures of ${P}_c$ and $\tilde{P}_c$ can be rather intuitive: for those who choose to venture into a new topic, how many of them are going to produce better research works than they used to, and for those who choose to stay in the same field, how many of them are able to keep the level of the scientific output? Our results demonstrate a strong positive correlation between $J$ and $P_c$, as well as between $\tilde{J}$ and $\tilde{P}_c$, where the Pearson correlation coefficient \textit{r} 0.78 for $J$ and 0.92 for $\tilde{J}$ (Figs. \ref{fig2}a-b). The positive correlation indicates that those who leave the current research field are more likely to produce works of more impact than their previous work. More importantly, it also implies that the further one leaves the original area, the higher this likelihood will be. 

The $P_c$ and $\tilde{P}_c$ provide a binary check on whether or not an author's works are on average better than her previous ones. They do not, however, 

\noindent quantify the extent of the improvement or the decline. It could be the case, for example, that the gain is small but the loss is big associated with a certain direction change. For this reason, we measure the relative impact change $(\bar{c}_{2,f} - \bar{c}_{2,i})/\bar{c}_{2,i}$, which quantifies the extend that the impact of the later research differs from the early ones. We calculate $(\bar{c}_{2,f} - \bar{c}_{2,i})/\bar{c}_{2,i}$ for each author and get the average value $R_c$ and $\tilde{R}_c$ of scientists with similar research direction change $J$ and $\tilde{J}$, respectively (for few scientists whose $\bar{c}_{2,i}=0$, we assign a default value 2 to their relative impact change). We again observe a strong and positive correlation between $J$ and $R_c$, as well as $\tilde{J}$ and $\tilde{R}_c$ (Figs. \ref{fig2}c-d). Taken together, those who exhibit a larger change of research direction demonstrate a higher extend of impact improvement. The larger the change, the more impact the work would receive.

A scientist may move to a ``hot'' field where the publication's average citation is higher than others. To explore if this can explain the positive correlation observed, we calculate the field-normalized citation of each paper (Section \hyperlink{Note.S3}{S3}). The positive correlation preserves when taking the difference of each field into account. Finally, to make sure our results are not affected by the sample sizes in calculating citations, we perform similar analyses using different citation measures. Besides, we also test the case when the citation time window is 3 years. The corresponding results are presented in Section \hyperlink{Note.S4}{S4}. In all cases, a strong positive correlation is observed between the research direction change and the impact change.

\subsection{Correlation between the direction change and productivity change}
Another quantity associated with scientific performance is productivity, typically quantified by the time needed to complete a certain number of papers. Here we use the time interval $t$ between the publications of the first and $m^{th}$ paper in a given series of papers. Similar to the above analyses, we identify the fraction of scientists $P_t$ (and $\tilde{P}_t$) whose productivity are increased ($t_f < t_i$), and calculate the average change rate $R_t$ (and $\tilde{R}_t$) of scientists' productivity by averaging $(\frac{1}{t_f} - \frac{1}{t_i})/\frac{1}{t_i}$ of individual scientists whose research direction change are within the same range (Figs. \ref{fig1}b-c).

\begin{figure}[H]
	\begin{center}
		\includegraphics[width=13.5cm]{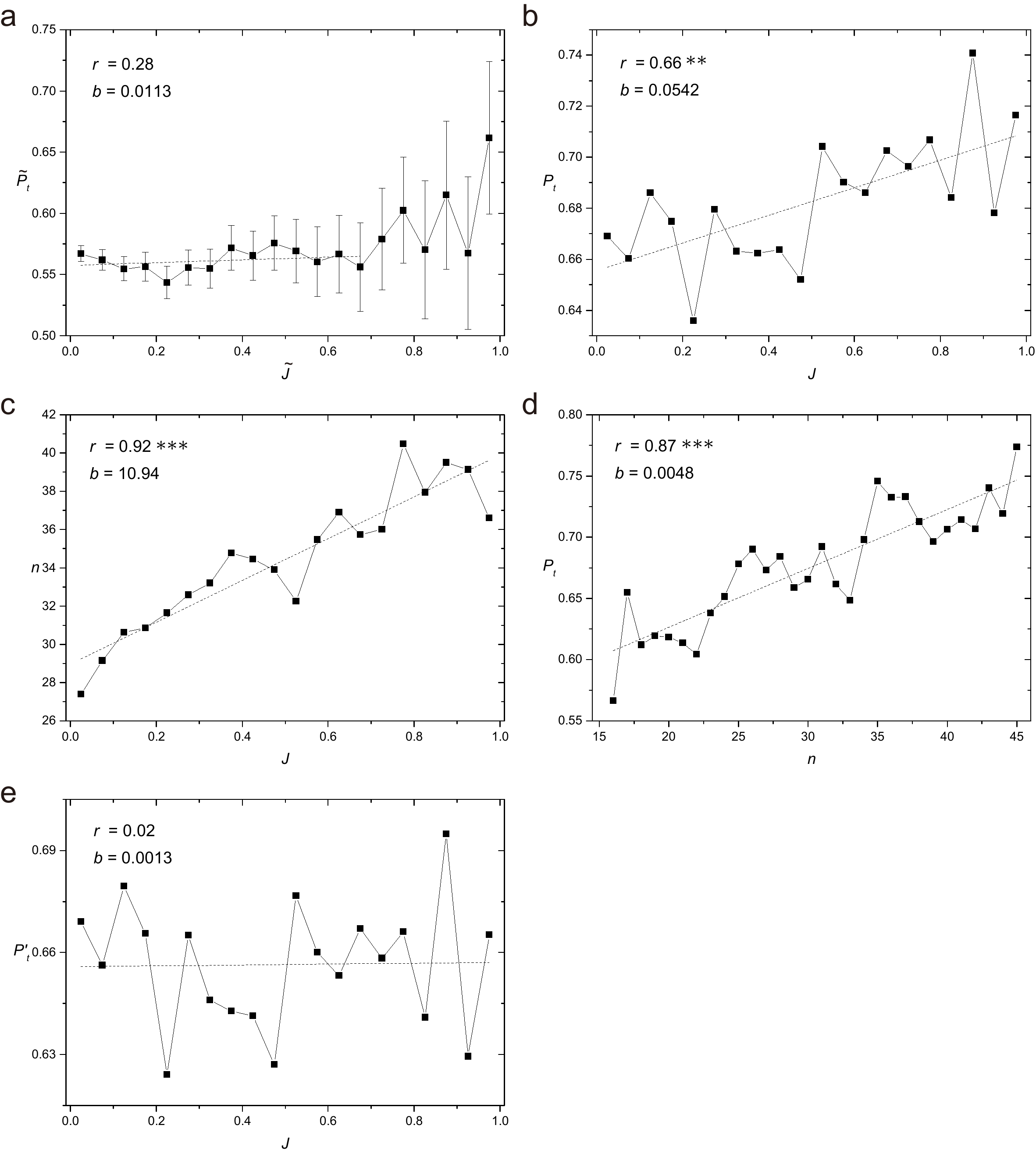}
		\caption{
			({\bf a}) $\tilde{P}_t$ is not correlated with $\tilde{J}$ for a range of values ($0 \leq \tilde{J} \leq 0.725$) with over 97\% of the sample size and small standard deviations. Due to the relatively small sample size (no more than 100) and high standard deviation for each group of $J$ in the range ($0.725 < \tilde{J} \leq 1.0$), we do not take this range into discussion.
			({\bf b}) $P_t$ increases with $J$, the slope of which is almost the same as the one predicted by the correlations between $P_t$ and $n$ as well as $n$ and $J$ (Section \protect\hyperlink{Note.S6}{S6}).
			({\bf c}) The average output $n$ conditioning on the range of direction change ($J-0.025,J+0.025$] is positively correlated with $J$.
			({\bf d}) $P_t$ is positively correlated with the output $n$.
			({\bf e}) After subtracting the increase induced by the pairwise dependence between $n$ and $J$ as well as $n$ and $P_t$, the result indicates that $P^{'}_t$ and $J$ are uncorrelated. The value of $b$ is defined as the slope of the corresponding linear regression function (The dashed line). *** $p < 0.001$, ** $p < 0.05$, * $p < 0.1$ ($t$-test for Pearson coefficient $r$).
			\label{fig3}}
	\end{center}
\end{figure}

The direct correlation measure shows two seemingly contradictory pictures. On one hand, based on two adjacent sets of papers, the research direction change $\tilde{J}$ is not correlated with the probability to increase productivity $\tilde{P}_t$ (Fig. \ref{fig3}a), nor the change of productivity rate $\tilde{R}_t$ (Fig. \hyperlink{figS3}{S3}a). On the other hand, if the change is measured based on two sets of papers that are at the two ends of a career, $J$ is positively correlated with both $P_t$ (Fig. \ref{fig3}b) and $R_t$ (Fig. \hyperlink{figS3}{S3}b). Does it mean that switching to new topics at the latter career would associate with advanced productivity?

To have a right understanding of the correlation between $J$ and $P_t$ as well as $J$ and $R_t$, we have to first control their inherent correlation. As $J$ depends on the two sets of papers that are maximally separated in one's publication list, the more publications one produces, the less likely it that the publications on one end of the list would contain similar topics with those on the other end. Therefore, it is expected and also empirically confirmed that $J$ is positively correlated with the number of publications $n$ in a career (Fig. \ref{fig3}c). Furthermore, the growing number of publications may enrich a scientist's experiences, skills, and collaboration networks, which in return would benefit productivity. It can be expected that the chance to surpass the publication rate at the beginning of the career (which usually corresponds to the graduate training period) would grow as one's publication list gets longer. Indeed, we observe empirically that $P_t$ and $R_t$ are positively correlated with $n$ (Fig. \ref{fig3}d, and Fig. \hyperlink{figS3}{S3}c).

The positive correlation between $J$ and $n$ as well as $P_t$ and $n$ leads to an inherent correlation between $J$ and $P_t$. In other words, given the way that $J$ is measured and the various values of the hidden variable $n$, $J$ is expected to correlate with $P_t$. Indeed, if we control the number of papers by focusing on $16 \le n \le 20$ or $21 \le n \le 25$, the correlation disappears (Section \hyperlink{Note.S5}{S5}).

To control this inherent correlation, we can first identify the dependence between $J$ and $P_t$ from their pairwise dependence with $n$ (Section \hyperlink{Note.S6}{S6}). The result gives $P_t \sim b\times J$, which is the expected dependence between $J$ and $P_t$ (red line in Fig. \ref{fig3}b). We then subtract the increase from $P_t$, yielding a measure $P^{'}_t = P_t - b\times J$ in which $P_t$'s natural dependence with $J$ is excluded. When the inherent correlation is properly controlled, we find that $P^{'}_t$ and $J$ are uncorrelated (Fig. \ref{fig3}e). The same analyses can also be performed on $R_t$, which leads to similar results (Fig. \hyperlink{figS3}{S3}d).

Taken together, by properly control the effect of publication number $n$ in our measure, we reach consistent results that the research direction change is not correlated with the probability to increase productivity nor the average change rate of productivity. This is contrary to our common perception that changing research direction is likely to hurt productivity. But we have to keep in mind that there may be a survivor bias. A scientist can contribute to the statistics only when she successfully changes the direction. Those who take the risk to explore a new area but fail to publish enough are not included in our data. Therefore, the interpretation of the results needs more caution. Finally, the hidden variable $n$ leads us to two contradictory relationships because it strongly and positively correlates with $P_t$ as well as $R_t$. In terms of impact change analyzed in the above section, we find that $n$ is weakly and slightly negatively correlated with $P_c$ and $R_c$ (Fig. \hyperlink{figS4}{S4}). This is in line with some previous findings that simply publishing a lot does not necessarily enhance the scientific impact of individual papers \citep{hanssen2015value,lariviere2016many,sarewitz2016pressure,kolesnikov2018researchers}. Therefore, although the dependence between $J$ and $n$ preserves, it does not play a role in our conclusion about impact change.

\subsection{Extended discussions}
It is a conventional narrative that scientific productivity tends to grow rapidly at the early stage of a career and then slowly declines. Such narrative is highlighted in some recent studies \citep{sinatra2016quantifying, li2020scientific} explaining why elite scientists tend to produce the most significant works at their early career stage: because their productivity declines when they get older. This pattern seems to be against our finding that $P_t$ and $n$ are positively correlated, which plays a crucial role in explaining the discrepancy in the initial results of the productivity change. Indeed, if we are not only interested in elite scientists, the productivity change can be diverse and the jump-decline pattern is not the only typical one \citep{way2017misleading}. Even in the work that implies a jump-decline pattern, we can still observe in the vast majority that the productivity declines at the very early career (typical schooling and training period) followed by a steady increase as the career unfolds (Fig. 1E in \cite{sinatra2016quantifying}). For a typical scientist, it may hold that the productivity at the end of the career, though has declined from the peak, is still higher than when she starts the career. Indeed, when we turn to a more comprehensive data set by the Web of Science, we still observe $P_t$ ($R_t$) and $n$ are positively correlated (Figs. \hyperlink{figS5}{S5}a-b). Another potentially important factor is the career length, which is not controlled in the above analyses. Therefore, those who are on the rising stage may outnumber those in the gloaming. Nevertheless, when the career length is controlled, the same positive correlation remains (Figs. \hyperlink{figS5}{S5}c-d). Moreover, the study by \cite{sinatra2016quantifying} and \cite{liu2018hot} suggest that the work with the highest impact may appear randomly in one's publication sequence. But this pattern is not contradictory to the positive correlation between the direction change and impact change, which implies that the work with the highest impact may appear in the later career. The ``random impact rule'' by \cite{sinatra2016quantifying, liu2018hot} is based on all samples without any control, whereas the correlation measured is conditioned on a certain research direction change.

Another work we wish to discuss more is the recent advances by \cite{zeng2019increasing} which also utilizes the APS data to quantify topic change patterns of scientists. In this pioneering work, the authors measure the probability to switch between topics in an individual career. The results demonstrate that compared with typical scientists, the top 10\% of productive scientists have a lower switching probability in the early career and higher switching probability in the later career. Moreover, the top 10\% most cited scientists have an overall lower switching probability than normal scientists. The relationship discovered seems contradictory to our findings. However, as a matter of fact, our work and the work by \cite{zeng2019increasing} are different in terms of the quantification of the direction change and the quality that the change is associated with. Hence, the two conclusions can be both valid. 

\begin{figure}[H]
	\begin{center}
		\includegraphics[width=13.5cm]{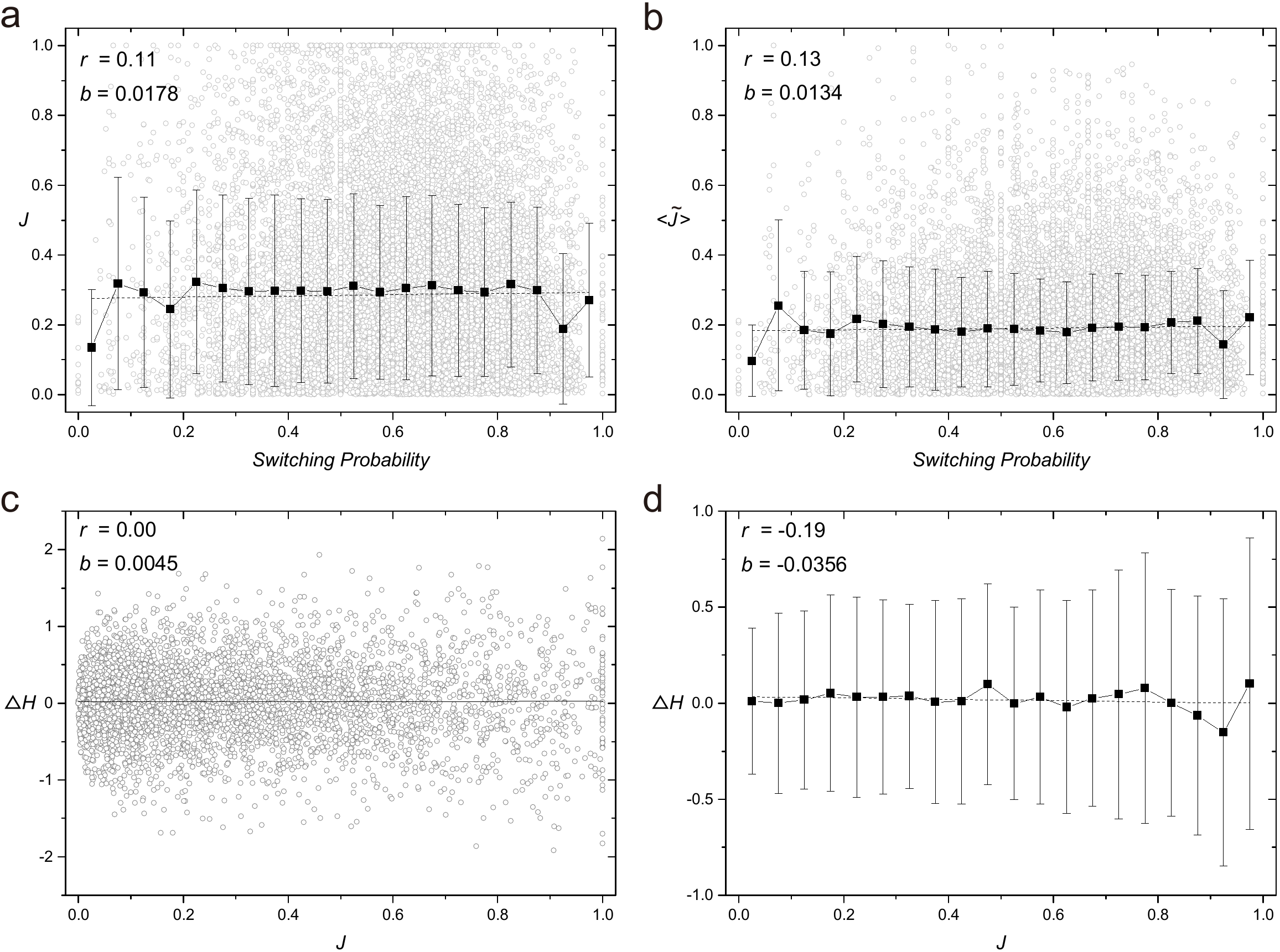}
		\caption{
			({\bf a}) For each scientist, we plot her $J$ versus switching probability (grey circle), and the mean value of $J$ conditioning on the range of (switching probability - 0.025, switching probability + 0.025] (scatter with line). The result shows that switching probability is not correlated with $J$ on the individual level ($p>0.1$).
			({\bf b}) For each scientist, we calculate the average value $\langle \tilde{J} \rangle$ of her $n-1$ $\tilde{J}$. Then we plot her $\langle \tilde{J} \rangle$ versus switching probability (grey circle), and the mean value of $\langle \tilde{J} \rangle$ conditioning on the range of (switching probability - 0.025, switching probability + 0.025] (scatter with line). The result shows that switching probability is not correlated with $\langle \tilde{J} \rangle$ at the individual level ($p>0.1$). 
			({\bf c}) For scientists whose impact has increased ($\bar{c}_{2,f} > \bar{c}_{2,i}$), we plot their $J$ versus the change of Shannon entropy $\Delta H$ (grey circle). The result shows that $\Delta H$ and $J$ are not correlated ($p>0.1$). The average $\Delta H$ is close to 0, indicating that diversity change is not associated with the impact increase.
			({\bf d}) Similar to ({\bf c}), but the mean value of $\Delta H$ is taken for individuals with ($J-0.025,J+0.025$].
			The value of $b$ is defined as the slope of the corresponding linear regression function (The dashed line). Error bars represent the one standard deviation of the mean.
			\label{fig4}}
	\end{center}
\end{figure}

First, in this work, we are interested in the change of performance, not the performance itself. In other words, we investigate how likely a scientist can publish works with more impact or within less time as the career develops, regardless if she is in the top 10\% elite group or not. The focus, and consequently the findings are different from that in \cite{zeng2019increasing}. More importantly, the switching probability quantifies the likelihood that the current topic would be different from the subsequent one, which only depends on the topic of two papers. The direction change applied in this study compares the averaged topics of two sets of papers. Therefore, they reflect two distinct aspects in an individual's research topic selection. As an example, assume a scientist who has published 16 papers within which the first 8 papers are on topic A and the next 8 are on topic B. The direction change $J$ of this scientist would be 1, as she explores totally different topics in the early and late sets of publications. The switching probability, however, is very low because this scientist only switches the topic once, which occurs in the 9th paper. Likewise, assume another scientist who has also published 16 papers on either topic A or topic B. She publishes following a pattern that if her current publication is on topic A, then the next paper will be on topic B, and vice versa. In this case, the direction change $J$ would be 0 as this scientist equally devotes herself to topics A and B throughout the career. But the switching probability would be very high as she constantly switches the topic. The two examples vividly depict the difference between these two measures. Indeed, when we try to associate a scientist's direction change and switching probability together, we find that they two are independent of each other (Figs. \ref{fig4}a-b). Therefore, we believe that this work and the work by \cite{zeng2019increasing} are complementary rather than contradictory. As they reveal two essential patterns underlying the scientific careers of individuals, it would be interesting to check in future studies if a combination of they two is a more comprehensive predictor for an individual career. 

Finally, we also wish to point out that our finding is different from that on research diversity, though they both suggest a positive citation gain \citep{2015Does,2015Are,2021Exploring,leahey2017prominent}. Indeed, the direction change and diversity change are two different things. Take the above example again, a scientist who has published the first 8 papers on topic A and the next 8 papers on topic B. The direction change is 1. But the diversity change is 0, as she always focuses on one topic only. Another example is when a scientist only focuses on topic A at the beginning and then changes the agenda with 80\% of topic A and five other topics each with 4\% of weight. As the main focus is still on topic A, the direction is very small. But because there are five other topics added to the agenda, the diversity change is high. Here, we use Shannon entropy to quantify the diversity of one's research agenda as $H =\sum^{67}_{j=1} - x_{j}log(x_{j})$ where $x_{j}$ is the $j^{th}$ element of the vector $g$ (the other two types of diversity index are reported in Section \hyperlink{Note.S7}{S7}). We focus on scientists whose impact has increased ($\bar{c}_{2,f} > \bar{c}_{2,i}$) and plot their change of entropy $\Delta H$ with their change of research direction. We find these two quantities are not correlated (Figs. \ref{fig4}c-d and Fig. \hyperlink{figS15}{S15}). The fact that the change of diversity is close to 0 on average indicates that the increased impact is not from a more interdisciplinary or more narrow research agenda. Instead, it is likely a result of bringing existing expertise to the new field and solving the field problem by a new approach.

\section{Conclusions}

To summarize, by utilizing PACS codes to classify topics in physics publications, we quantify the degree of the research direction change in an individual career. Instead of considering the overall scientific performance, we associate this direction change of an individual with her change of performance. On one hand, we find that the direction change is strongly and positively correlated with the change of impact. Those who demonstrate a larger change on research topics are more likely to receive more citations than they used to. The magnitude of the relative improvement also tends to be higher. On the other hand, the direction change is not correlated with the productivity change. The likelihood that one increases or decreases productivity may be associated with many factors in an individual career, but the change of research direction alone is not associated with it. We perform supplementary analyses to demonstrate the robustness of the conclusion. We also discuss in detail the difference between our findings and others'. Our study provides another point of view not fully captured in previous studies.

The statistics provide an encouraging prediction for scientists who venture into a new field. Once they are established in the new field, they are likely to become better scientists. What is not mentioned, however, is the risk associated with the direction change \citep{bromham2016interdisciplinary, azoulay2011incentives, foster2015tradition, goldstein2020know}. Indeed, our analyses are based on scientists who have published enough papers in the new field. Those who try to change but fail to have the new research published are not included. This naturally poses a survival bias in the result, which motivates further studies on failures across a career \citep{yin2019quantifying, wang2019early}. There are also many factors that the correlation measure alone can not explain. For example, a scientist may be forced to leave the old field because it shrinks and can not yield interesting results, and another scientist finds a way to combine her prior knowledge with the new problem which leads to fruitful results \citep{pramanik2019migration}. In both cases, a citation gain and a direction change would be observed. Yet, in the current study we can not distinguish them nor identify which causes the other. Given confounding factors in an individual career, it is difficult but important to check other explanations of the observation. 
Moreover, our study is based on APS publication data. Despite the intensive usage \citep{liu2017knowledge, chinazzi2019mapping}, this data set only covers a small part of the science. The classification scheme provided by the PACS code has certain limitations as well. Given the availability of large scale data sets, such as Microsoft Academic Graph \citep{wang2020microsoft} and advances in machine learning tools to identify and classify topics from papers \citep{qian2020understanding, chinazzi2019mapping, palmucci2020your, shen2019node2vec}, it would be important to check if similar patterns can be observed in other data set. Finally, we adopt the whole counting in this study that gives equal credit to all co-authors. This approach may lead to inflation when calculating the impact of a scientist's work and her productivity \citep{huang2011counting,yu2021papers}. It may be worthwhile to try other counting or credit allocation method \citep{sivertsen2019measuring,shen2014collective,wang2019nonlinear}.

\section*{Supplementary Information}

See the file of supplementary information for additional materials.

\section*{Acknowledgments}

This work is supported by the National Natural Science Foundation of China (No. 61603309). Boleslaw K. Szymanski is also supported by the Army Research Office (ARO) under Grant W911NF-16-1-0524.
\newpage

\bibliographystyle{apalike2}
\biboptions{authoryear}
\bibliography{arxiv}

\begin{thebibliography}{}

\bibitem[Aleta et~al., 2019]{aleta2019explore}
Aleta, A., Meloni, S., Perra, N., \& Moreno, Y. (2019).
\newblock Explore with caution: mapping the evolution of scientific interest in
  physics.
\newblock {\em EPJ Data Science}, 8(1), 1--15.

\bibitem[Alfredo et~al., 2015]{2015Does}
Alfredo, Y.~Y., Ismael, R., Pablo, D., \& Wolfgang, G. (2015).
\newblock Does interdisciplinary research lead to higher citation impact? the
  different effect of proximal and distal interdisciplinarity.
\newblock {\em Plos One}, 10(8), e0135095.

\bibitem[AlShebli et~al., 2018]{alshebli2018preeminence}
AlShebli, B.~K., Rahwan, T., \& Woon, W.~L. (2018).
\newblock The preeminence of ethnic diversity in scientific collaboration.
\newblock {\em Nature communications}, 9(1), 5163.

\bibitem[Amjad et~al., 2018]{amjad2018measuring}
Amjad, T., Daud, A., \& Song, M. (2018).
\newblock Measuring the impact of topic drift in scholarly networks.
\newblock In {\em Companion Proceedings of the The Web Conference 2018}  (pp.\
  373--378).

\bibitem[Azoulay et~al., 2011]{azoulay2011incentives}
Azoulay, P., Graff~Zivin, J.~S., \& Manso, G. (2011).
\newblock Incentives and creativity: evidence from the academic life sciences.
\newblock {\em The RAND Journal of Economics}, 42(3), 527--554.

\bibitem[Bromham et~al., 2016]{bromham2016interdisciplinary}
Bromham, L., Dinnage, R., \& Hua, X. (2016).
\newblock Interdisciplinary research has consistently lower funding success.
\newblock {\em Nature}, 534(7609), 684--687.

\bibitem[Bu et~al., 2018]{bu2018understanding}
Bu, Y., Ding, Y., Xu, J., Liang, X., Gao, G., \& Zhao, Y. (2018).
\newblock Understanding success through the diversity of collaborators and the
  milestone of career.
\newblock {\em Journal of the Association for Information Science and
  Technology}, 69(1), 87--97.

\bibitem[Chen et~al., 2015]{2015Are}
Chen, S., Arsenault, C., \& Larivière, V. (2015).
\newblock Are top-cited papers more interdisciplinary?
\newblock {\em Journal of Informetrics}, 9(4), 1034--1046.

\bibitem[Chen et~al., 2021a]{chen2021exploring}
Chen, S., Qiu, J., Arsenault, C., \& Larivi{\`e}re, V. (2021a).
\newblock Exploring the interdisciplinarity patterns of highly cited papers.
\newblock {\em Journal of Informetrics}, 15(1), 101124.

\bibitem[Chen et~al., 2021b]{2021Exploring}
Chen, S., Qiu, J., Arsenault, C., \& Larivière, V. (2021b).
\newblock Exploring the interdisciplinarity patterns of highly cited papers.
\newblock {\em Journal of Informetrics}, 15(1), 101124.

\bibitem[Chen et~al., 2021c]{chen2020rank}
Chen, W., Zhu, Z., \& Jia, T. (2021c).
\newblock The rank boost by inconsistency in university rankings: evidence from
  14 rankings of chinese universities.
\newblock {\em Quantitative Science Studies}, 2(1), 335--349.

\bibitem[Chinazzi et~al., 2019]{chinazzi2019mapping}
Chinazzi, M., Gon{\c{c}}alves, B., Zhang, Q., \& Vespignani, A. (2019).
\newblock Mapping the physics research space: a machine learning approach.
\newblock {\em EPJ Data Science}, 8(1), 1--18.

\bibitem[Clauset et~al., 2015]{clauset2015systematic}
Clauset, A., Arbesman, S., \& Larremore, D.~B. (2015).
\newblock Systematic inequality and hierarchy in faculty hiring networks.
\newblock {\em Science advances}, 1(1), e1400005.

\bibitem[Deville et~al., 2014]{deville2014career}
Deville, P., Wang, D., Sinatra, R., Song, C., Blondel, V.~D., \& Barab{\'a}si,
  A.-L. (2014).
\newblock Career on the move: Geography, stratification, and scientific impact.
\newblock {\em Scientific reports}, 4(1), 4770.

\bibitem[Foster et~al., 2015]{foster2015tradition}
Foster, J.~G., Rzhetsky, A., \& Evans, J.~A. (2015).
\newblock Tradition and innovation in scientists’ research strategies.
\newblock {\em American Sociological Review}, 80(5), 875--908.

\bibitem[Goldstein \& Kearney, 2020]{goldstein2020know}
Goldstein, A.~P. \& Kearney, M. (2020).
\newblock Know when to fold ‘em: An empirical description of risk management
  in public research funding.
\newblock {\em Research Policy}, 49(1), 103873.

\bibitem[Hanssen \& J{\o}rgensen, 2015]{hanssen2015value}
Hanssen, T.-E.~S. \& J{\o}rgensen, F. (2015).
\newblock The value of experience in research.
\newblock {\em Journal of Informetrics}, 9(1), 16--24.

\bibitem[Hu et~al., 2020]{hu2020describing}
Hu, X., Li, X., \& Rousseau, R. (2020).
\newblock Describing citations as a function of time.
\newblock {\em Journal of Data and Information Science}, 5(2), 1--12.

\bibitem[Huang et~al., 2020a]{huang2020comparison}
Huang, C.-K., Neylon, C., Brookes-Kenworthy, C., Hosking, R., Montgomery, L.,
  Wilson, K., \& Ozaygen, A. (2020a).
\newblock Comparison of bibliographic data sources: Implications for the
  robustness of university rankings.
\newblock {\em Quantitative Science Studies}, 1(2), 445--478.

\bibitem[Huang et~al., 2020b]{huang2020historical}
Huang, J., Gates, A.~J., Sinatra, R., \& Barab{\'a}si, A.-L. (2020b).
\newblock Historical comparison of gender inequality in scientific careers
  across countries and disciplines.
\newblock {\em Proceedings of the National Academy of Sciences}, 117(9),
  4609--4616.

\bibitem[Huang et~al., 2011]{huang2011counting}
Huang, M.-H., Lin, C.-S., \& Chen, D.-Z. (2011).
\newblock Counting methods, country rank changes, and counting inflation in the
  assessment of national research productivity and impact.
\newblock {\em Journal of the American society for information science and
  technology}, 62(12), 2427--2436.

\bibitem[Huang et~al., 2020c]{huang2020patent}
Huang, Y., Chen, L., \& Zhang, L. (2020c).
\newblock Patent citation inflation: The phenomenon, its measurement, and
  relative indicators to temper its effects.
\newblock {\em Journal of Informetrics}, 14(2), 101015.

\bibitem[Jia et~al., 2017]{jia2017quantifying}
Jia, T., Wang, D., \& Szymanski, B.~K. (2017).
\newblock Quantifying patterns of research-interest evolution.
\newblock {\em Nature Human Behaviour}, 1(4), 0078.

\bibitem[Jones \& Weinberg, 2011]{jones2011age}
Jones, B.~F. \& Weinberg, B.~A. (2011).
\newblock Age dynamics in scientific creativity.
\newblock {\em Proceedings of the National Academy of Sciences}, 108(47),
  18910--18914.

\bibitem[King, 2004]{king2004scientific}
King, D.~A. (2004).
\newblock The scientific impact of nations.
\newblock {\em Nature}, 430(6997), 311--316.

\bibitem[Kolesnikov et~al., 2018]{kolesnikov2018researchers}
Kolesnikov, S., Fukumoto, E., \& Bozeman, B. (2018).
\newblock Researchers’ risk-smoothing publication strategies: Is productivity
  the enemy of impact?
\newblock {\em Scientometrics}, 116(3), 1995--2017.

\bibitem[{Kuhn}, 1977]{kuhn1977the}
{Kuhn}, T.~S. (1977).
\newblock The essential tension : selected studies in scientific tradition and
  change.
\newblock {\em Journal for the Scientific Study of Religion}, 18(3), 328.

\bibitem[Larivi{\`e}re \& Costas, 2016]{lariviere2016many}
Larivi{\`e}re, V. \& Costas, R. (2016).
\newblock How many is too many? on the relationship between research
  productivity and impact.
\newblock {\em PloS one}, 11(9), e0162709.

\bibitem[Leahey et~al., 2017]{leahey2017prominent}
Leahey, E., Beckman, C.~M., \& Stanko, T.~L. (2017).
\newblock Prominent but less productive: The impact of interdisciplinarity on
  scientists’ research.
\newblock {\em Administrative Science Quarterly}, 62(1), 105--139.

\bibitem[Li et~al., 2020]{li2020scientific}
Li, J., Yin, Y., Fortunato, S., \& Wang, D. (2020).
\newblock Scientific elite revisited: patterns of productivity, collaboration,
  authorship and impact.
\newblock {\em Journal of the Royal Society Interface}, 17(165), 20200135.

\bibitem[Liu et~al., 2018]{liu2018hot}
Liu, L., Wang, Y., Sinatra, R., Giles, C.~L., Song, C., \& Wang, D. (2018).
\newblock Hot streaks in artistic, cultural, and scientific careers.
\newblock {\em Nature}, 559(7714), 396--399.

\bibitem[Liu et~al., 2020]{liu2020dominance}
Liu, L., Yu, J., Huang, J., Xia, F., \& Jia, T. (2020).
\newblock The dominance of big teams in china’s scientific output.
\newblock {\em Quantitative Science Studies}, 2(1), 350--362.

\bibitem[Liu et~al., 2017]{liu2017knowledge}
Liu, W., Nanetti, A., \& Cheong, S.~A. (2017).
\newblock Knowledge evolution in physics research: An analysis of bibliographic
  coupling networks.
\newblock {\em Plos one}, 12(9), e0184821.

\bibitem[Ma et~al., 2020]{ma2020mentorship}
Ma, Y., Mukherjee, S., \& Uzzi, B. (2020).
\newblock Mentorship and prot{\'e}g{\'e} success in stem fields.
\newblock {\em Proceedings of the National Academy of Sciences}, 117(25),
  14077--14083.

\bibitem[Maule{\'o}n \& Bordons, 2006]{mauleon2006productivity}
Maule{\'o}n, E. \& Bordons, M. (2006).
\newblock Productivity, impact and publication habits by gender in the area of
  materials science.
\newblock {\em Scientometrics}, 66(1), 199--218.

\bibitem[Mukherjee et~al., 2017]{mukherjee2017nearly}
Mukherjee, S., Romero, D.~M., Jones, B., \& Uzzi, B. (2017).
\newblock The nearly universal link between the age of past knowledge and
  tomorrow’s breakthroughs in science and technology: The hotspot.
\newblock {\em Science advances}, 3(4), e1601315.

\bibitem[Palmucci et~al., 2020]{palmucci2020your}
Palmucci, A., Liao, H., Napoletano, A., \& Zaccaria, A. (2020).
\newblock Where is your field going? a machine learning approach to study the
  relative motion of the domains of physics.
\newblock {\em PloS one}, 15(6), e0233997.

\bibitem[Petersen, 2018]{petersen2018multiscale}
Petersen, A.~M. (2018).
\newblock Multiscale impact of researcher mobility.
\newblock {\em Journal of The Royal Society Interface}, 15(146), 20180580.

\bibitem[Petersen et~al., 2011]{petersen2011quantitative}
Petersen, A.~M., Jung, W.-S., Yang, J.-S., \& Stanley, H.~E. (2011).
\newblock Quantitative and empirical demonstration of the matthew effect in a
  study of career longevity.
\newblock {\em Proceedings of the National Academy of Sciences}, 108(1),
  18--23.

\bibitem[Petersen et~al., 2019]{petersen2019methods}
Petersen, A.~M., Pan, R.~K., Pammolli, F., \& Fortunato, S. (2019).
\newblock Methods to account for citation inflation in research evaluation.
\newblock {\em Research Policy}, 48(7), 1855--1865.

\bibitem[Pramanik et~al., 2019]{pramanik2019migration}
Pramanik, S., Gora, S.~T., Sundaram, R., Ganguly, N., \& Mitra, B. (2019).
\newblock On the migration of researchers across scientific domains.
\newblock In {\em Proceedings of the International AAAI Conference on Web and
  Social Media}, volume~13  (pp.\ 381--392).

\bibitem[Qian et~al., 2020]{qian2020understanding}
Qian, Y., Liu, Y., \& Sheng, Q.~Z. (2020).
\newblock Understanding hierarchical structural evolution in a scientific
  discipline: A case study of artificial intelligence.
\newblock {\em Journal of Informetrics}, 14(3), 101047.

\bibitem[Radicchi et~al., 2008]{radicchi2008universality}
Radicchi, F., Fortunato, S., \& Castellano, C. (2008).
\newblock Universality of citation distributions: Toward an objective measure
  of scientific impact.
\newblock {\em Proceedings of the National Academy of Sciences}, 105(45),
  17268--17272.

\bibitem[Robinson-Garcia et~al., 2019]{robinson2019many}
Robinson-Garcia, N., Sugimoto, C.~R., Murray, D., Yegros-Yegros, A.,
  Larivi{\`e}re, V., \& Costas, R. (2019).
\newblock The many faces of mobility: Using bibliometric data to measure the
  movement of scientists.
\newblock {\em Journal of Informetrics}, 13(1), 50--63.

\bibitem[Sarewitz, 2016]{sarewitz2016pressure}
Sarewitz, D. (2016).
\newblock The pressure to publish pushes down quality.
\newblock {\em Nature}, 533(7602), 147--147.

\bibitem[Shen \& Barab{\'a}si, 2014]{shen2014collective}
Shen, H.-W. \& Barab{\'a}si, A.-L. (2014).
\newblock Collective credit allocation in science.
\newblock {\em Proceedings of the National Academy of Sciences}, 111(34),
  12325--12330.

\bibitem[Shen et~al., 2019]{shen2019node2vec}
Shen, Z., Chen, F., Yang, L., \& Wu, J. (2019).
\newblock Node2vec representation for clustering journals and as a possible
  measure of diversity.
\newblock {\em Journal of Data and Information Science}, 4(2), 79--92.

\bibitem[Sinatra et~al., 2016]{sinatra2016quantifying}
Sinatra, R., Wang, D., Deville, P., Song, C., \& Barab{\'a}si, A.-L. (2016).
\newblock Quantifying the evolution of individual scientific impact.
\newblock {\em Science}, 354(6312), aaf5239.

\bibitem[Sivertsen et~al., 2019]{sivertsen2019measuring}
Sivertsen, G., Rousseau, R., \& Zhang, L. (2019).
\newblock Measuring scientific contributions with modified fractional counting.
\newblock {\em Journal of Informetrics}, 13(2), 679--694.

\bibitem[Wang et~al., 2013]{wang2013quantifying}
Wang, D., Song, C., \& Barab{\'a}si, A.-L. (2013).
\newblock Quantifying long-term scientific impact.
\newblock {\em Science}, 342(6154), 127--132.

\bibitem[Wang et~al., 2019a]{wang2019nonlinear}
Wang, F., Fan, Y., Zeng, A., \& Di, Z. (2019a).
\newblock A nonlinear collective credit allocation in scientific publications.
\newblock {\em Scientometrics}, 119(3), 1655--1668.

\bibitem[Wang et~al., 2015]{wang2015interdisciplinarity}
Wang, J., Thijs, B., \& Gl{\"a}nzel, W. (2015).
\newblock Interdisciplinarity and impact: Distinct effects of variety, balance,
  and disparity.
\newblock {\em PloS one}, 10(5), e0127298.

\bibitem[Wang et~al., 2020]{wang2020microsoft}
Wang, K., Shen, Z., Huang, C., Wu, C.-H., Dong, Y., \& Kanakia, A. (2020).
\newblock Microsoft academic graph: When experts are not enough.
\newblock {\em Quantitative Science Studies}, 1(1), 396--413.

\bibitem[Wang et~al., 2019b]{wang2019early}
Wang, Y., Jones, B.~F., \& Wang, D. (2019b).
\newblock Early-career setback and future career impact.
\newblock {\em Nature communications}, 10(1), 4331.

\bibitem[Way et~al., 2017]{way2017misleading}
Way, S.~F., Morgan, A.~C., Clauset, A., \& Larremore, D.~B. (2017).
\newblock The misleading narrative of the canonical faculty productivity
  trajectory.
\newblock {\em Proceedings of the National Academy of Sciences}, 114(44),
  E9216--E9223.

\bibitem[Way et~al., 2019]{way2019productivity}
Way, S.~F., Morgan, A.~C., Larremore, D.~B., \& Clauset, A. (2019).
\newblock Productivity, prominence, and the effects of academic environment.
\newblock {\em Proceedings of the National Academy of Sciences}, 116(22),
  10729--10733.

\bibitem[Wu et~al., 2019]{wu2019large}
Wu, L., Wang, D., \& Evans, J.~A. (2019).
\newblock Large teams develop and small teams disrupt science and technology.
\newblock {\em Nature}, 566(7744), 378--382.

\bibitem[Yin et~al., 2019]{yin2019quantifying}
Yin, Y., Wang, Y., Evans, J.~A., \& Wang, D. (2019).
\newblock Quantifying the dynamics of failure across science, startups and
  security.
\newblock {\em Nature}, 575(7781), 190--194.

\bibitem[Yu et~al., 2021]{yu2021papers}
Yu, J., Yin, C., Liu, L., \& Jia, T. (2021).
\newblock A paper's corresponding affiliation and first affiliation are
  consistent at the country level in web of science.
\newblock {\em arXiv e-prints}, arXiv:2101.09426.

\bibitem[Zeng et~al., 2019]{zeng2019increasing}
Zeng, A., Shen, Z., Zhou, J., Fan, Y., Di, Z., Wang, Y., Stanley, H.~E., \&
  Havlin, S. (2019).
\newblock Increasing trend of scientists to switch between topics.
\newblock {\em Nature communications}, 10(1), 3439.

\bibitem[Zhang et~al., 2017]{zhang2017identifying}
Zhang, C., Liu, C., Yu, L., Zhang, Z.-K., \& Zhou, T. (2017).
\newblock Identifying the academic rising stars via pairwise citation increment
  ranking.
\newblock In {\em Asia-Pacific Web (APWeb) and Web-Age Information Management
  (WAIM) Joint Conference on Web and Big Data}  (pp.\ 475--483).

\bibitem[Zhao et~al., 2020]{zhao2020investigation}
Zhao, Z., Bu, Y., Kang, L., Min, C., Bian, Y., Tang, L., \& Li, J. (2020).
\newblock An investigation of the relationship between scientists’ mobility
  to/from china and their research performance.
\newblock {\em Journal of Informetrics}, 14(2), 101037.

\bibitem[Zuo \& Zhao, 2018]{zuo2018more}
Zuo, Z. \& Zhao, K. (2018).
\newblock The more multidisciplinary the better?--the prevalence and
  interdisciplinarity of research collaborations in multidisciplinary
  institutions.
\newblock {\em Journal of Informetrics}, 12(3), 736--756.

\end{thebibliography}

\newpage

\section*{Supplementary Information}
%
\begin{figure}[h]
	\begin{center}
		\hypertarget{figS1}{\includegraphics[width=13.5cm]{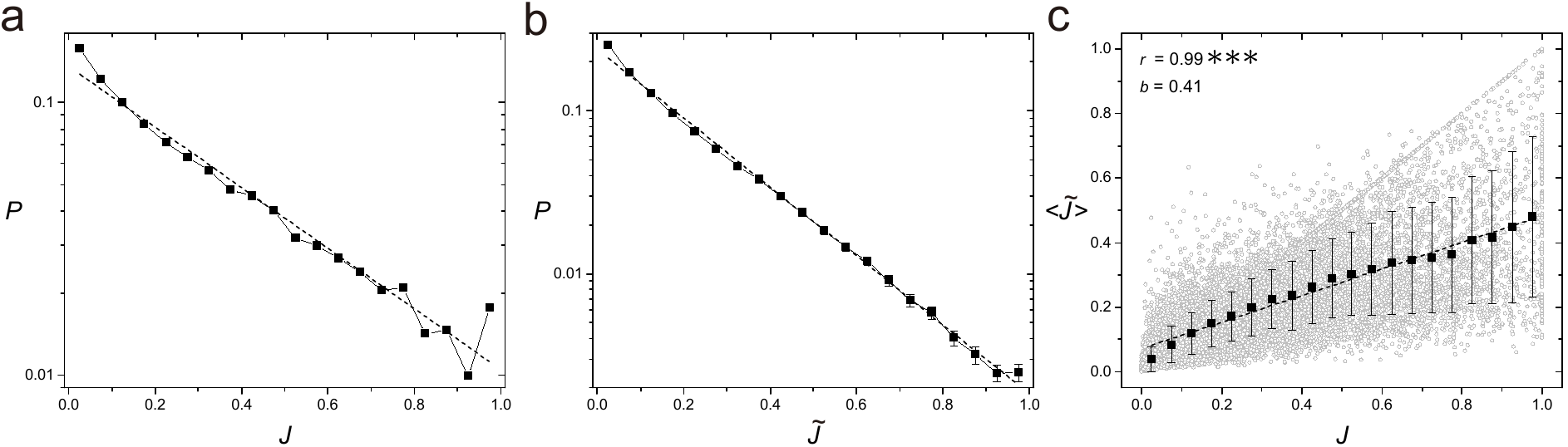}}
		\caption*{Figure S1: 
			({\bf a}) The fraction of scientists $P$ within a range of ($J-0.025,J+0.025$] drops exponentially with $J$.
			({\bf b}) The fraction of scientists $\tilde{P}$ within a range of ($\tilde{J} - 0.025, \tilde{J} + 0.025$] drops exponentially with $\tilde{J}$.
			({\bf c}) For a scientist with $n$ papers, we calculate the average value $\langle \tilde{J} \rangle$ of her $n-1$ $\tilde{J}$. Then we plot her $\langle \tilde{J} \rangle$ versus $J$ (grey circle), and the mean value of $\langle \tilde{J} \rangle$ conditioning on the range of ($J-0.025,J+0.025$] (scatter with line). The result shows that $J$ and $\tilde{J}$ are consistent at the individual level. The value of $b$ is defined as the slope of the corresponding linear regression function (The dashed line). *** $p < 0.001$, ** $p < 0.05$, * $p < 0.1$ ($t$-test for Pearson coefficient $r$). Error bars represent the one standard deviation of the mean.}
	\end{center}
\end{figure}\noindent
\clearpage
\begin{figure}[h]
	\begin{center}
		\hypertarget{figS2}{\includegraphics[width=13.5cm]{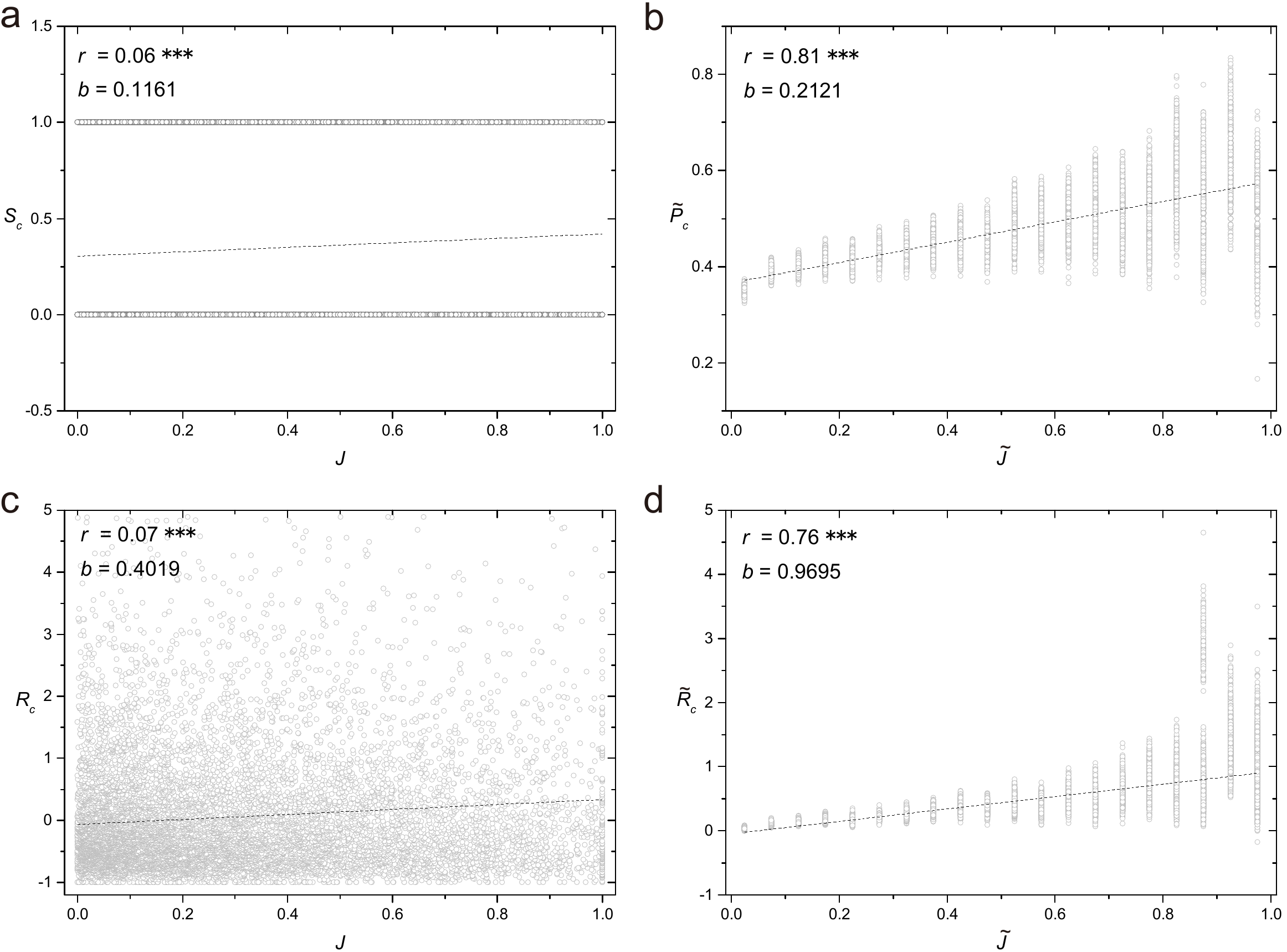}}
		\caption*{Figure S2: 
			({\bf a}) Each grey circle of the scatter plot corresponds to a scientist, where $S_c$ represents the state of impact change ($S_c=1$ means that a scientist has increased her impact, and $S_c=0$ means not). The correlation coefficient $r$ between $J$ and $S_c$ in the scatter plot is 0.06 ($p<0.001$), indicating that the correlation between $J$ and $S_c$ is strongly significant.			
			({\bf b}) The correlation coefficient $r$ between $\tilde{J}$ and $\tilde{P}_c$ in the scatter plot is 0.81.
			({\bf c}) Each grey circle of the scatter plot corresponds to a scientist. The correlation coefficient $r$ between $J$ and $R_c$ in the scatter plot is 0.07 ($p<0.001$), indicating that the correlation between $J$ and $R_c$ is strongly significant.
			({\bf d}) The correlation coefficient $r$ between $\tilde{J}$ and $\tilde{R}_c$ in the scatter plot is 0.76. The value of $b$ is defined as the slope of the corresponding linear regression function (The dashed line). *** $p < 0.001$, ** $p < 0.05$, * $p < 0.1$ ($t$-test for Pearson coefficient $r$).}
	\end{center}
\end{figure}\noindent
\clearpage
\begin{figure}[h]
	\begin{center}
		\hypertarget{figS3}{\includegraphics[width=13.5cm]{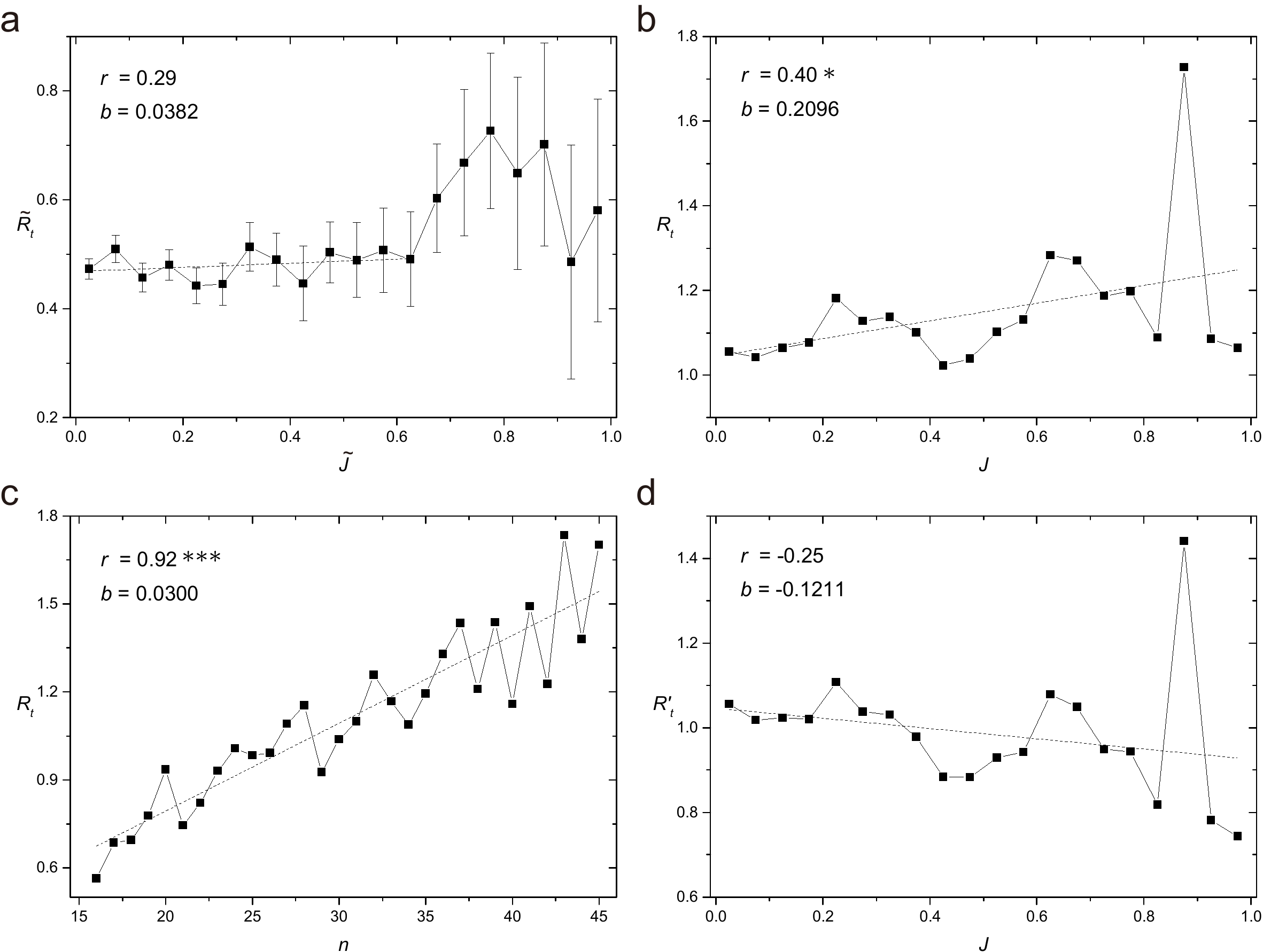}}
		\caption*{Figure S3:
			({\bf a}) $\tilde{R}_t$ is not correlated with $\tilde{J}$ for a range of values ($0 \leq \tilde{J} \leq 0.675$) with over 96\% of the sample size and small standard deviations. Due to the relatively small sample size (no more than 100) and high standard deviation for each group of $J$ in the range ($0.675 < \tilde{J} \leq 1.0$), we do not take this range into discussion.
			({\bf b}) $R_t$ has a weak increase trend with $J$, the slope of which is smaller than the one predicted by the correlations between $R_t$ and $n$ as well as $n$ and $J$ (Section \hyperlink{Note.S6}{S6}).
			({\bf c}) $R_t$ is positively correlated to the output $n$.
			({\bf d}) After subtracting the increase induced by the pairwise dependence between $n$ and $J$ as well as $n$ and $R_t$ (Section \hyperlink{Note.S6}{S6}), the result indicates that $R^{'}_t$ and $J$ are uncorrelated. The value of $b$ is defined as the slope of the corresponding linear regression function (The dashed line). *** $p < 0.001$, ** $p < 0.05$, * $p < 0.1$ ($t$-test for Pearson coefficient $r$).}
	\end{center}
\end{figure}\noindent
\clearpage
\begin{figure}[h]
	\begin{center}
		\hypertarget{figS4}{\includegraphics[width=13.5cm]{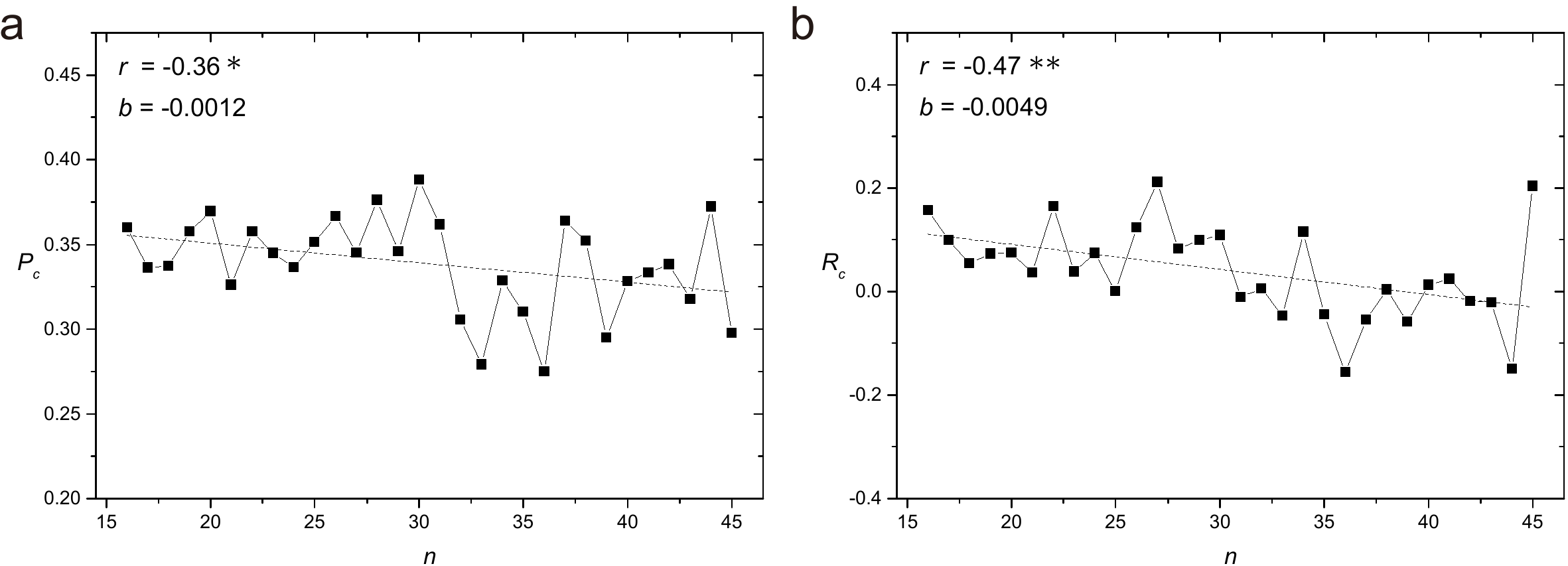}}
		\caption*{Figure S4:
			({\bf a-b}) In the data of APS analyzed in main text, $P_c$ and $R_c$ have a weak negative correlation with output $n$, basing on $c_2$. The value of $b$ is defined as the slope of the corresponding linear regression function (The dashed line). *** $p < 0.001$, ** $p < 0.05$, * $p < 0.1$ ($t$-test for Pearson coefficient $r$).}
	\end{center}
\end{figure}\noindent
\clearpage
\begin{figure}[h]
	\begin{center}
		\hypertarget{figS5}{\includegraphics[width=13.5cm]{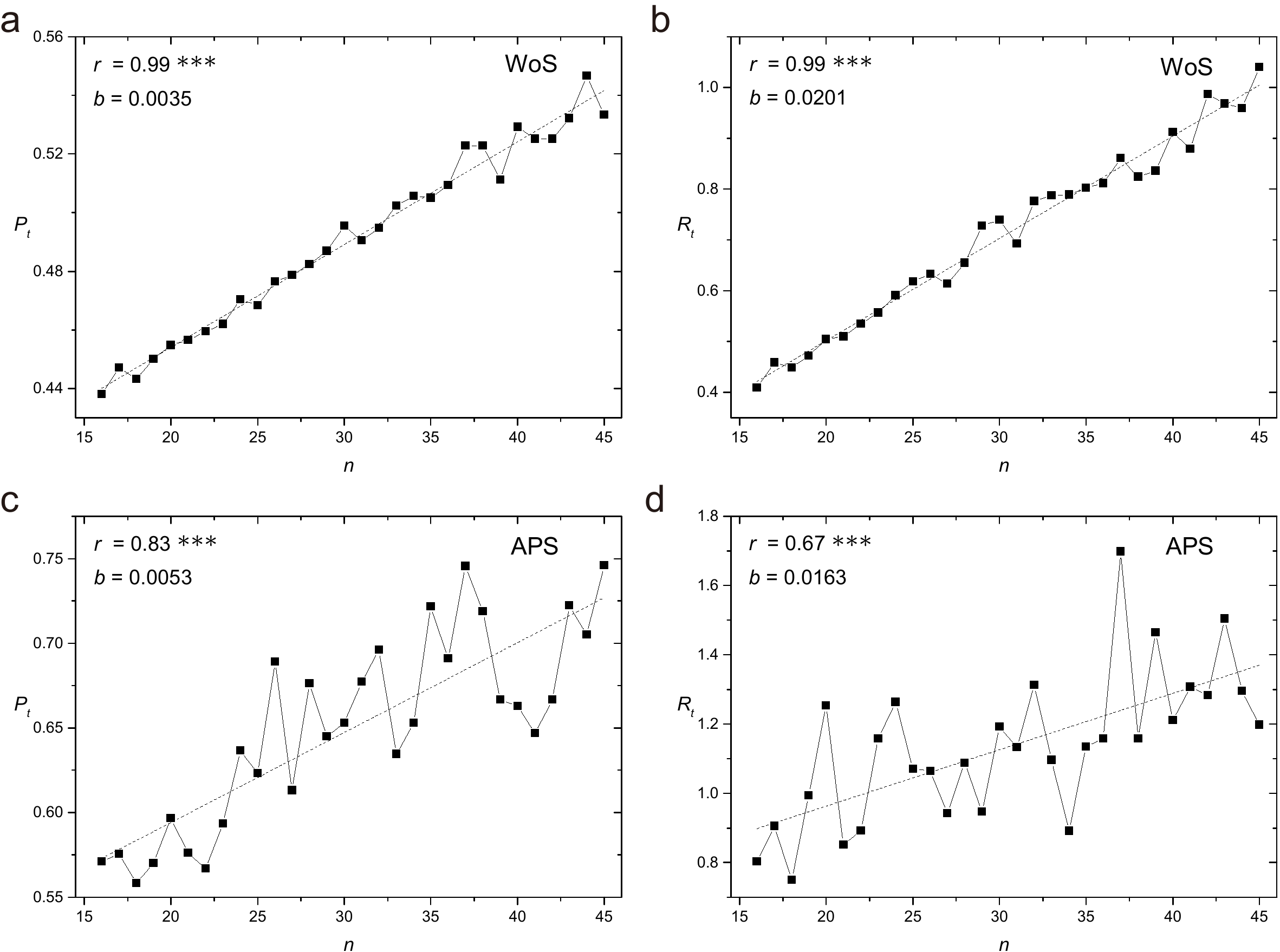}}
		\caption*{Figure S5:
			({\bf a-b}) In the data of Web of Science (WoS), $P_t$ and $R_t$ are positively correlated with the output $n$.
			({\bf c-d}) In the data of APS analyzed in main text, we condition on authors whose career length is no less than 20 years, and find that $P_t$ and $R_t$ are positively correlated with the output $n$. The value of $b$ is defined as the slope of the corresponding linear regression function (The dashed line). *** $p < 0.001$, ** $p < 0.05$, * $p < 0.1$ ($t$-test for Pearson coefficient $r$).}
	\end{center}
\end{figure}\noindent
\clearpage

\hypertarget{Note.S1}{\noindent{\bf S1. Citation normalization}}
\\

We consider the raw citations received within 2 years after publishing, $c_{2,raw}$, as a publication's scientific impact. To compare the impact of papers published at different time periods, we normalize the $c_{2,raw}$ of each publication by the average $\langle c_{2,raw}\rangle$ of the publishing year and multiplying by 10, as $c_{2} = \frac{c_{2,raw}}{\langle c_{2,raw}\rangle}*10$ utilized in the main text, where 10 is an arbitrary constant that has no quantitative effect to our investigations but restores the citation quantity $c_{2}$ to a relatively realistic value. Hence, the consequent $c_{2}$ provides a rational and comparable measure of scientific impact across years.
\clearpage

\hypertarget{Note.S2}{\noindent{\bf S2. Results based on the choices of $m$}}
\\

As $m = 8$ is taken into account in our analysis in the main text, we also investigate other cases when $m = 7$ and $m = 9$. Overall, our observation does not change with the value of $m$. The main observations based on $m = 7$ and $m = 9$ are shown in Fig. \hyperlink{figS6}{S6} and Fig. \hyperlink{figS7}{S7} respectively. There are 17,525 scientists with $n \geq 14$ and 12,368 scientists with $n \geq 18$, corresponding to the size of populations after excluding papers with a Shannon entropy $H>2.5$ in each analysis.
\begin{figure}[h]
	\begin{center}
		\hypertarget{figS6}{\includegraphics[width=13.5cm]{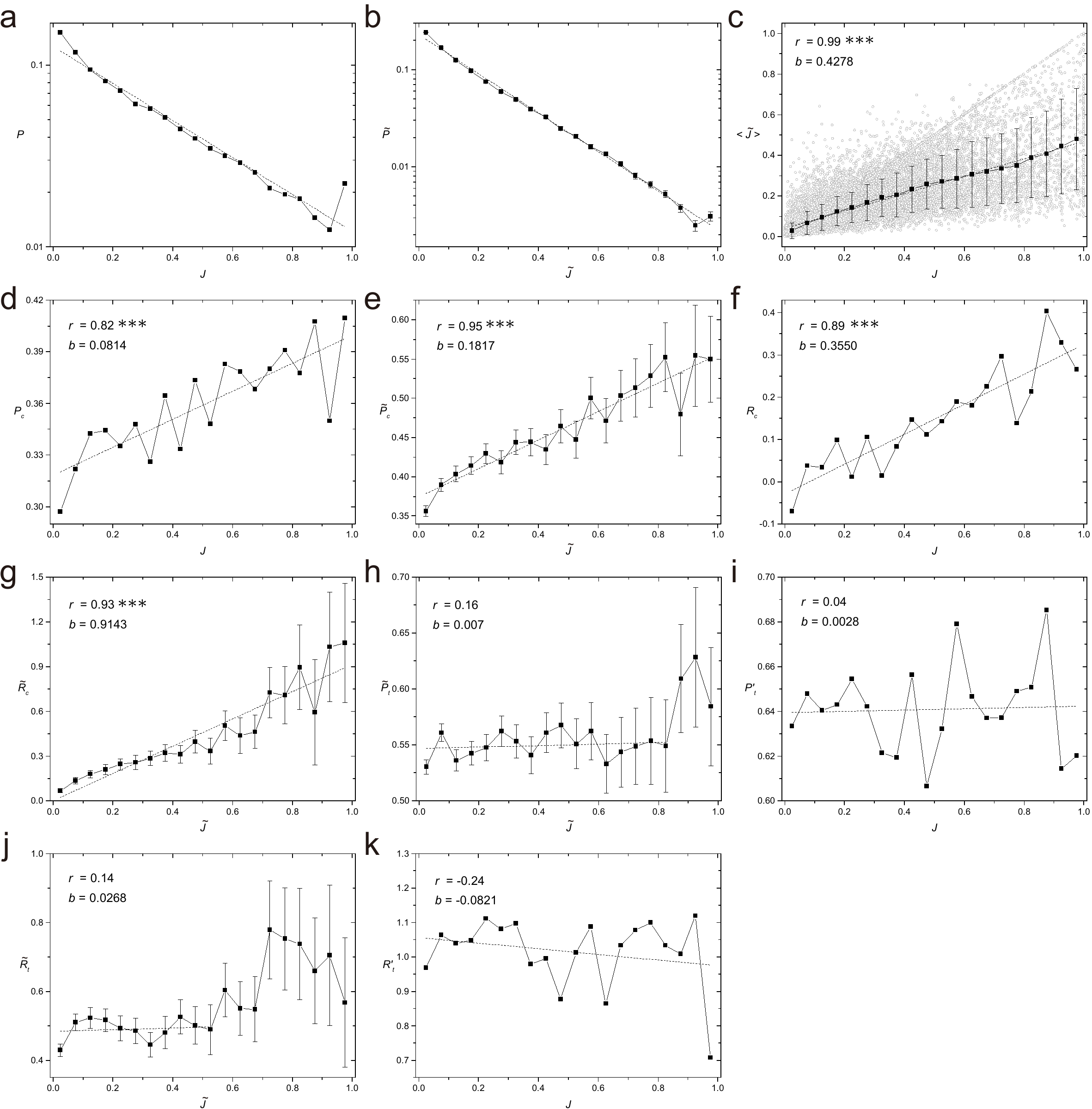}}
		\caption*{Figure S6:
			The results based on $m=7$. 
			({\bf a-b}) The fraction of scientists $P$ and $\tilde{P}$ drop exponentially with $J$ and $\tilde{J}$.
			({\bf c}) For a scientist with $n$ papers, we calculate the average value $\langle \tilde{J} \rangle$ of her $n-1$ $\tilde{J}$. Then we plot her $\langle \tilde{J} \rangle$ versus $J$ (grey circle), and the mean value of $\langle \tilde{J} \rangle$ conditioning on the range of ($J-0.025,J+0.025$] (scatter with line). The result shows that $J$ and $\tilde{J}$ are consistent at the individual level.
			({\bf d-e}) $P_c$ and $\tilde{P}_c$ are positively correlated with $J$ and $\tilde{J}$.
			({\bf f-g}) $R_c$ and $\tilde{R}_c$ are positively correlated with $J$ and $\tilde{J}$.
			({\bf h}) $\tilde{P}_t$ is not correlated with $\tilde{J}$ for a range of values ($0 \leq \tilde{J} \leq 0.875$) with over 99\% of the sample size and small standard deviations. Due to the relatively small sample size and high standard deviation for each group of $J$ in the range ($0.875 < \tilde{J} \leq 1.0$), we do not take this range into discussion.
			({\bf i}) After subtracting the increase induced by the pairwise dependence between $n$ and $J$ as well as $n$ and $P_t$ (Section \hyperlink{Note.S6}{S6}), the result indicates that $P^{'}_t$ and $J$ are uncorrelated.
			({\bf j}) $\tilde{R}_t$ is not correlated with $\tilde{J}$ for a range of values ($0 \leq \tilde{J} \leq 0.575$) with over 93\% of the sample size and small standard deviations. Due to the relatively small sample size and high standard deviation for each group of $J$ in the range ($0.575 < \tilde{J} \leq 1.0$), we do not take this range into discussion.
			({\bf k}) After subtracting the increase induced by the pairwise dependence between $n$ and $J$ as well as $n$ and $R_t$ (Section \hyperlink{Note.S6}{S6}), the result indicates that $R^{'}_t$ and $J$ are uncorrelated.
			The value of $b$ is defined as the slope of the corresponding linear regression function (The dashed line). *** $p < 0.001$, ** $p < 0.05$, * $p < 0.1$ ($t$-test for Pearson coefficient $r$).}
	\end{center}
\end{figure}\noindent
\begin{figure}[h]
	\begin{center}
		\hypertarget{figS7}{\includegraphics[width=13.5cm]{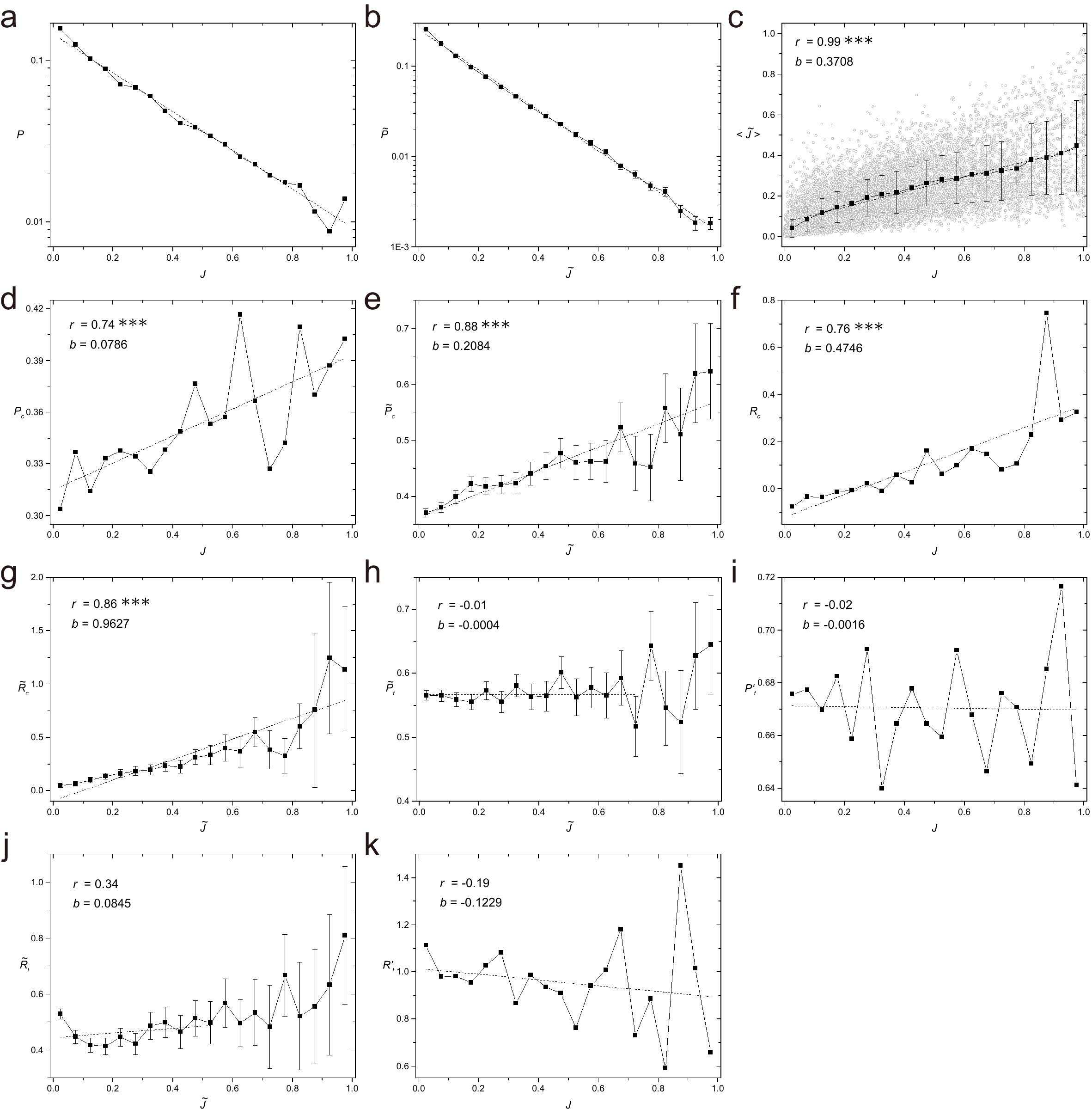}}
		\caption*{Figure S7:
			The results based on $m=9$. 
			({\bf a-b}) The fraction of scientists $P$ and $\tilde{P}$ drop exponentially with $J$ and $\tilde{J}$.
			({\bf c}) For a scientist with $n$ papers, we calculate the average value $\langle \tilde{J} \rangle$ of her $n-1$ $\tilde{J}$. Then we plot her $\langle \tilde{J} \rangle$ versus $J$ (grey circle), and the mean value of $\langle \tilde{J} \rangle$ conditioning on the range of ($J-0.025,J+0.025$] (scatter with line). The result shows that $J$ and $\tilde{J}$ are consistent at the individual level.
			({\bf d-e}) $P_c$ and $\tilde{P}_c$ are positively correlated with $J$ and $\tilde{J}$.
			({\bf f-g}) $R_c$ and $\tilde{R}_c$ are positively correlated with $J$ and $\tilde{J}$.
			({\bf h}) $\tilde{P}_t$ is not correlated with $\tilde{J}$ for a range of values ($0 \leq \tilde{J} \leq 0.775$) with over 98\% of the sample size and small standard deviations. Due to the relatively small sample size and high standard deviation for each group of $J$ in the range ($0.775 < \tilde{J} \leq 1.0$), we do not take this range into discussion.
			({\bf i}) After subtracting the increase induced by the pairwise dependence between $n$ and $J$ as well as $n$ and $P_t$ (Section \hyperlink{Note.S6}{S6}), the result indicates that $P^{'}_t$ and $J$ are uncorrelated.
			({\bf j}) $\tilde{R}_t$ is not correlated with $\tilde{J}$ for a range of values ($0 \leq \tilde{J} \leq 0.575$) with over 94\% of the sample size and small standard deviations. Due to the relatively small sample size and high standard deviation for each group of $J$ in the range ($0.575 < \tilde{J} \leq 1.0$), we do not take this range into discussion.
			({\bf k}) After subtracting the increase induced by the pairwise dependence between $n$ and $J$ as well as $n$ and $R_t$ (Section \hyperlink{Note.S6}{S6}), the result indicates that $R^{'}_t$ and $J$ are uncorrelated.
			The value of $b$ is defined as the slope of the corresponding linear regression function (The dashed line). *** $p < 0.001$, ** $p < 0.05$, * $p < 0.1$ ($t$-test for Pearson coefficient $r$).}
	\end{center}
\end{figure}\noindent
\clearpage

\hypertarget{Note.S3}{\noindent{\bf S3. Impact change based on field-normalized citation}}
\\

By allocating the citation $c_{2,raw}^p$ of a paper to each PACS code $i$ it contains through fractional counting method as $c_{2,raw}^i=c_{2,raw}^p*f_i$, where $f_i$ represents the fraction of PACS code $i$ in a paper, we obtain the yearly average citation $\langle c_{2,raw}^i\rangle$ of PACS code $i$, and find that the $\langle c_{2,raw}^i\rangle$ of different PACS codes can be very different (Fig. \hyperlink{figS8}{S8}). After obtaining the yearly average citation $\langle c_{2,raw}^i\rangle$ of each PACS code, we calculate the field citation of a paper by $c_{2,portfolio}^p=\sum_{i} \langle c_{2,raw}^i\rangle * f_i$, which represents a paper’s weighted citation as a combination of the yearly average citation of each PACS code it contains. We then normalize the $c_{2,raw}^p$ by $c_{2,portfolio}^p$ and multiplying by 10, as $c_2^p=\frac{c_{2,raw}^p}{c_{2,portfolio}^p}*10$. We perform the same analyses using the field-normalized citation and find the same patterns (Fig. \hyperlink{figS9}{S9}). Therefore, the increased impact is not because scientists tend to move to a hot field.
\begin{figure}[h]
	\begin{center}
		\hypertarget{figS8}{\includegraphics[width=10cm]{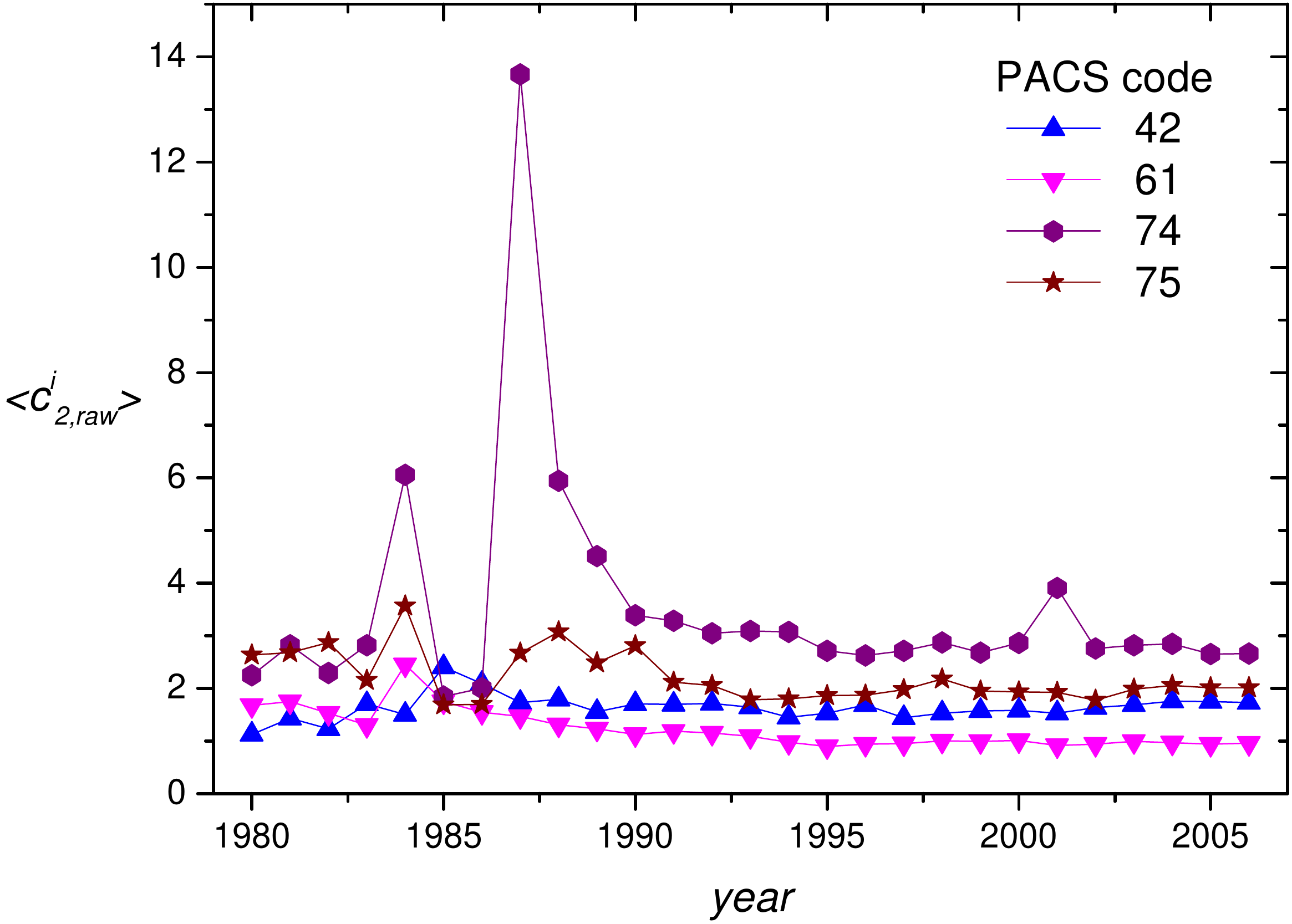}}
		\caption*{Figure S8:
			We allocate the citation to each PACS code of a paper by fractional counting to obtain the field-normalized citation of each paper. Here we depict four of the most focused PACS codes, and find that the yearly $\langle c_{2,raw}^i\rangle$ of each PACS code are different.}
	\end{center}
\end{figure}\noindent
\begin{figure}[h]
	\begin{center}
		\hypertarget{figS9}{\includegraphics[width=13.5cm]{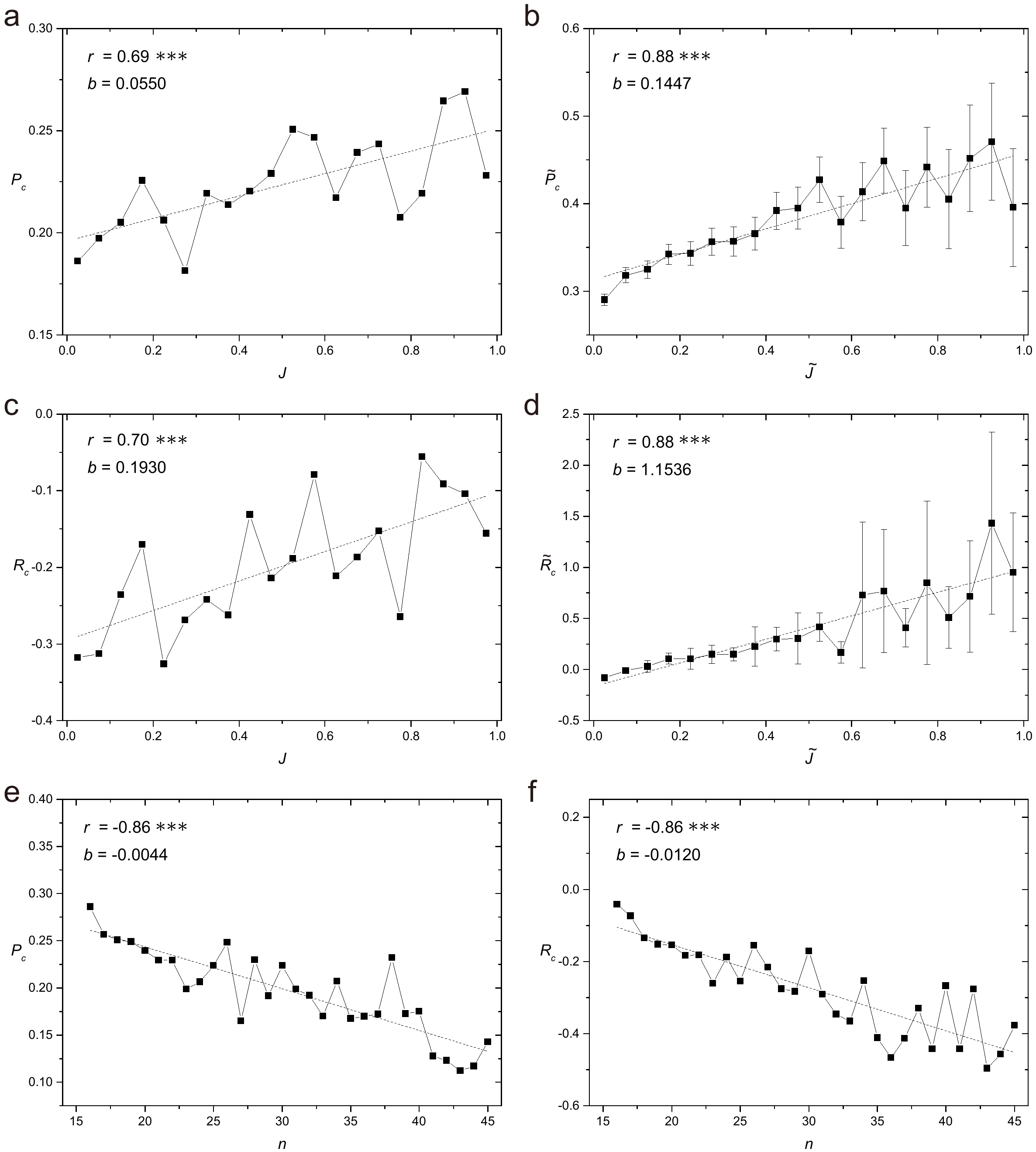}}
		\caption*{Figure S9:
			The results of the impact change according to the field-normalized citation, basing on $m=8$. ({\bf a-d}) The relationships between the impact change and direction change under the field-normalized citation are in line with results in the main text, as the positive correlations between $P_c$ ($R_c$) and $J$, as well as $\tilde{P}_c$ ($\tilde{R}_c$) and $J$ still hold. ({\bf e-f}) The negative correlation between $P_c$ ($R_c$) and output $n$ also remains. The value of $b$ is defined as the slope of the corresponding linear regression function (The dashed line). *** $p < 0.001$, ** $p < 0.05$, * $p < 0.1$ ($t$-test for Pearson coefficient $r$). Error bars represent the one standard deviation of the mean.}
	\end{center}
\end{figure}\noindent
\clearpage

\hypertarget{Note.S4}{\noindent{\bf S4. Measuring $\bar{c}_{2,i}$ and $\bar{c}_{2,f}$}}
\\

As the data set contains citation information by the year 2009, only papers published by 2007 receive the full $c_2$ after the publication. $\bar{c}_{2,i}$ and $\bar{c}_{2,f}$ are calculated based on the papers published by 2007 in the two paper sets. As scientists who have no qualified paper in the second paper set are excluded, the sample size shrinks from 14,726 originally to 13,170. It is possible for a scientist that the number of the qualified papers in the second paper set is less than $m$, but $m$ qualified papers in the first paper set, as papers are sequenced by the publication date. Hence $\bar{c}_{2,i}$ and $\bar{c}_{2,f}$ may be based on different sample sizes. In the relationship between $P_c$ ($R_c$) and $J$ in the main text, $\bar{c}_{2,i}$ is based on $m$ papers and $\bar{c}_{2,f}$ is based on the number of the qualified papers in the second paper set. To overcome the issue of different sample sizes, we take a different measurement in the relationship between $\tilde{P}_c$ ($\tilde{R}_c$) and $\tilde{J}$ in the main text. If the number $m'$ of the qualified papers in the second paper set is less than $m$, we randomly choose $m'$ papers in the first paper set to calculate $\bar{c}_{2,i}$.
\par To make sure our conclusion is not affected by the difference in the sample size, we apply some other variations as listed below. In all cases, positive correlations are observed.
\par 1. For the relationship between $P_c$ and $J$, if the number of the qualified papers in the second paper set ($m'$) is less than $m$ ($m = 8$), we randomly choose $m'$ papers in the first paper set to calculate $\bar{c}_{2,i}$, which is then used to compare with $\bar{c}_{2,f}$ to calculate $P_c$. As $P_c$ here depends on the choice of random samples for $\bar{c}_{2,i}$, we repeat the measurement 1000 times to obtain the mean and standard deviation. This is to make sure that the positive correlation is not affected by different sample sizes in calculating $\bar{c}_{2,i}$ and $\bar{c}_{2,f}$ . The result is shown in Fig. \hyperlink{figS10}{S10}a.
\par 2. To make sure the positive correlation between $\tilde{P}_c$ and $\tilde{J}$ is not affected by the small sample size for $\bar{c}_{2,f}$, we only consider the samples with the number of the qualified papers ($m'$) in the second paper set satisfying $m' \geq m/2$. The result is shown in Fig. \hyperlink{figS10}{S10}b.
\par 3. For the relationship between $P_c$ and $J$, we only consider the scientists who have no less than $2m$ papers by the year 2007 and calculate the $P_c$ and $J$ based on papers published by 2007. In this case, the sample size for $\bar{c}_{2,i}$ and $\bar{c}_{2,f}$ are the same. The result is shown in Fig. \hyperlink{figS10}{S10}c.
\par 4. To make sure the results are not affected by the cutoff year, we perform a similar measurement as that in 3. But we consider scientists who have equal or greater than $2m$ papers by the year 2006. $P_c$ and $J$ are calculated based on papers published by 2006. The result is shown in Fig. \hyperlink{figS10}{S10}d.
\par 5. Scientific impact based on $c_3$. Originally we count citations received within 2 years of the publications. To make sure our results are not affected by the choice of the 2-year time window, we also consider citations received within 3 years of the publications. Correspondingly we focus on scientists who have $n \geq 2m$ papers and whose last $m$ papers contain at least one paper published by 2006, and the qualified population shrinks to 11,584. We perform the same measurements as in the main text for this case and the results are shown in Figs. \hyperlink{figS10}{S10}e-f.
\par 6. We repeat the above 5 measurements for growth rate $R_c$ and $\tilde{R}_c$ (Fig. \hyperlink{figS11}{S11}).
\begin{figure}[h]
	\begin{center}
		\hypertarget{figS10}{\includegraphics[width=13.5cm]{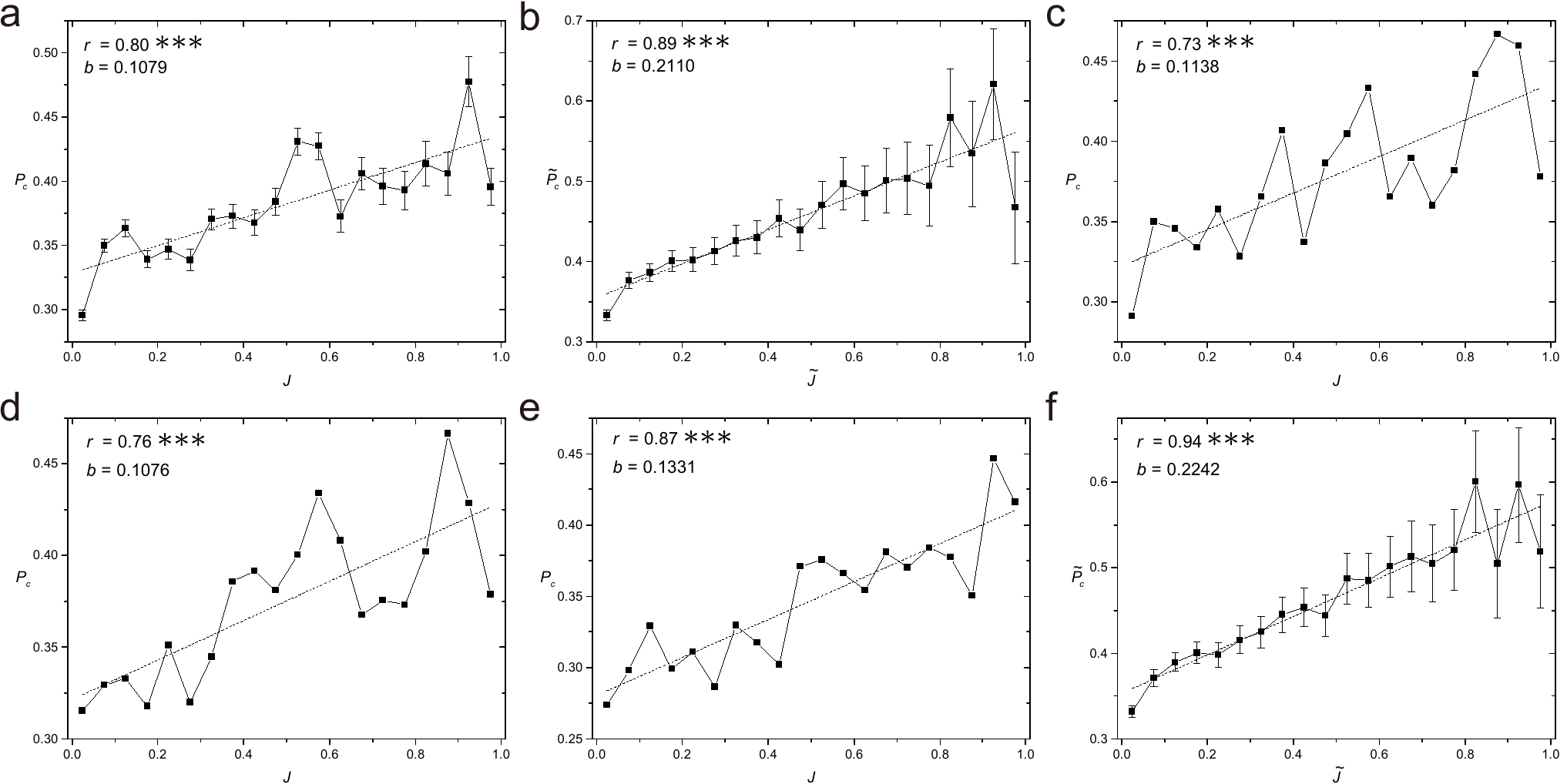}}
		\caption*{Figure S10:
			Different variations in calculating $P_c$ and $\tilde{P}_c$.}
	\end{center}
\end{figure}\noindent
\begin{figure}[h]
	\begin{center}
		\hypertarget{figS11}{\includegraphics[width=13.5cm]{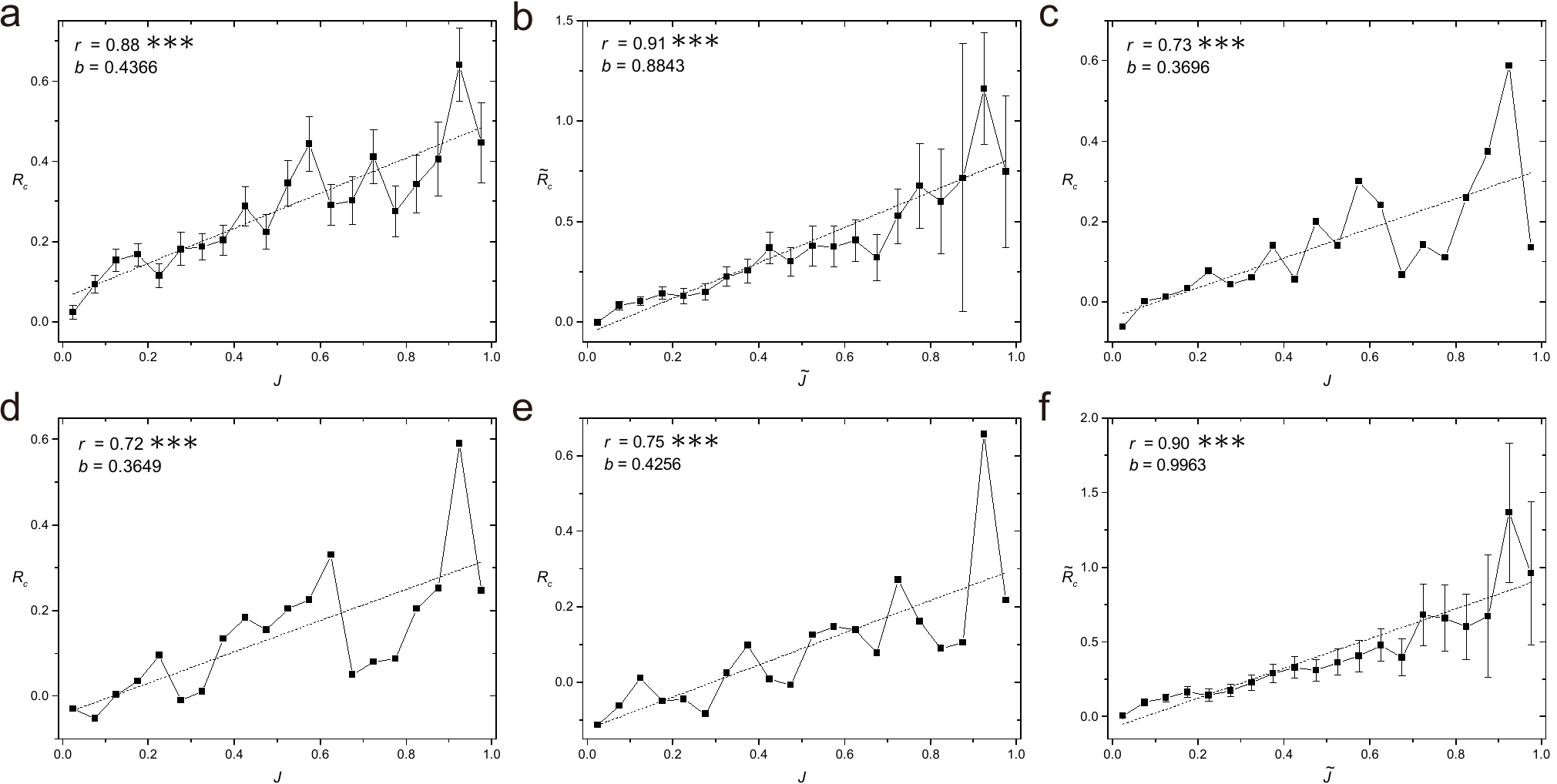}}
		\caption*{Figure S11:
			Different variations in calculating $R_c$ and $\tilde{R}_c$.}
	\end{center}
\end{figure}\noindent
\clearpage

\hypertarget{Note.S5}{\noindent{\bf S5. Results conditioning on the output $n$}}
\\

By sampling authors into the range $(n-2.5,n+2.5]$ of publication output $n$, we find the ranges (15, 20] and (20,25] contain the top 2 largest population, corresponding to groups of scientists who on average published 17.5 and 22.5 papers in her career (Fig. \hyperlink{figS12}{S12}). By conditioning on scientists whose outputs $n$ fall into the region (15, 20] and (20,25] respectively, we observe the flat relationship between $P_t$ (and $R_t$) and $J$, confirming our conclusions in the main text that the dependence of $P_t$ (and $R_t$) to $J$ is mainly driven by the hidden variable $n$ in the scenario of the measurement of $J$. (Figs. \hyperlink{figS13}{S13}-\hyperlink{figS14}{S14}).
\begin{figure}[h]
	\begin{center}
		\hypertarget{figS12}{\includegraphics[width=10cm]{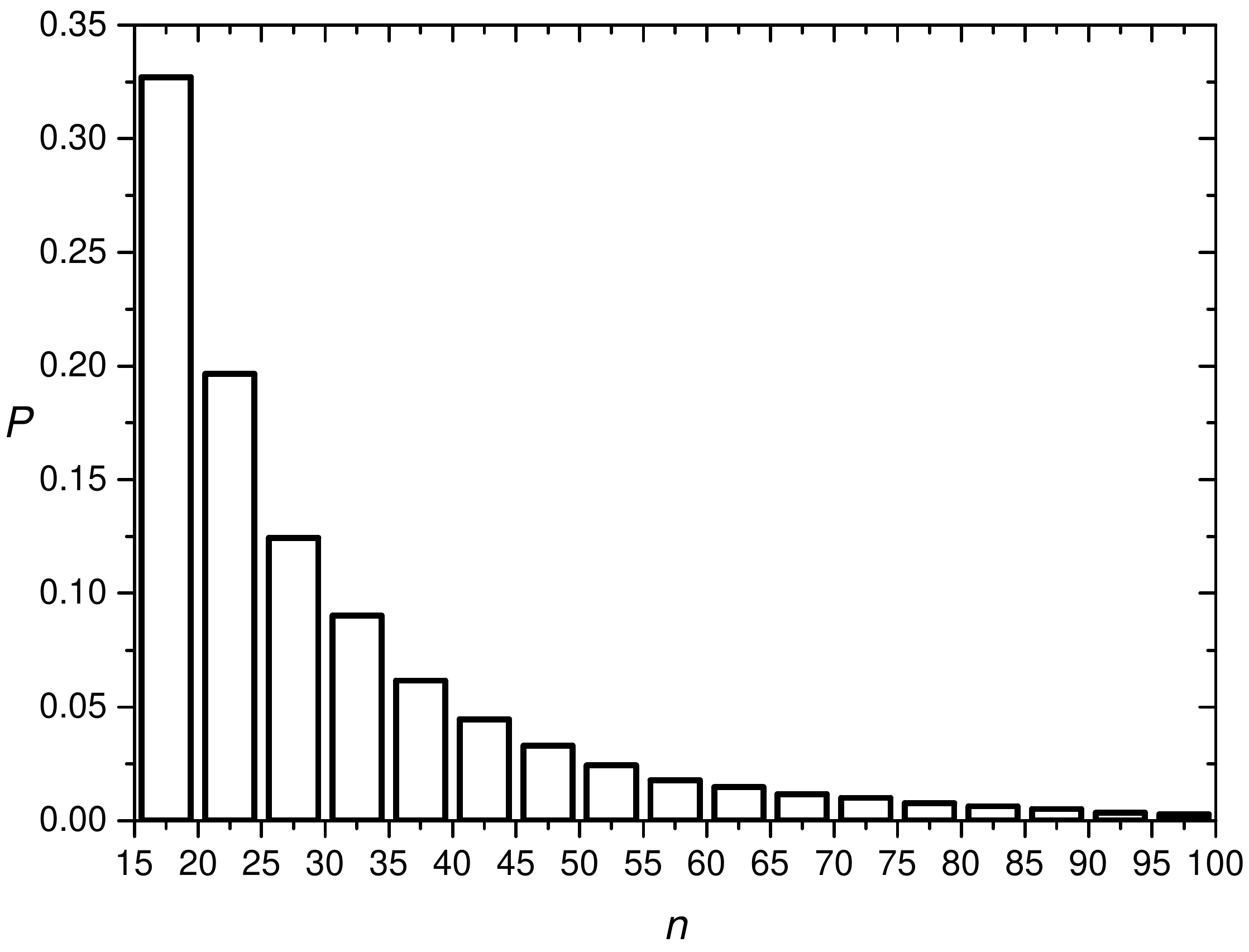}}
		\caption*{Figure S12:
			By sampling authors into the range $(n-2.5,n+2.5]$ of publication output $n$, we find the region (15, 20] and (20,25] contain the first and second-largest population of the $n$ distribution, corresponding to groups of scientists who on average publish 17.5 and 22.5 papers in her career.}
	\end{center}
\end{figure}\noindent
\begin{figure}[h]
	\begin{center}
		\hypertarget{figS13}{\includegraphics[width=13.5cm]{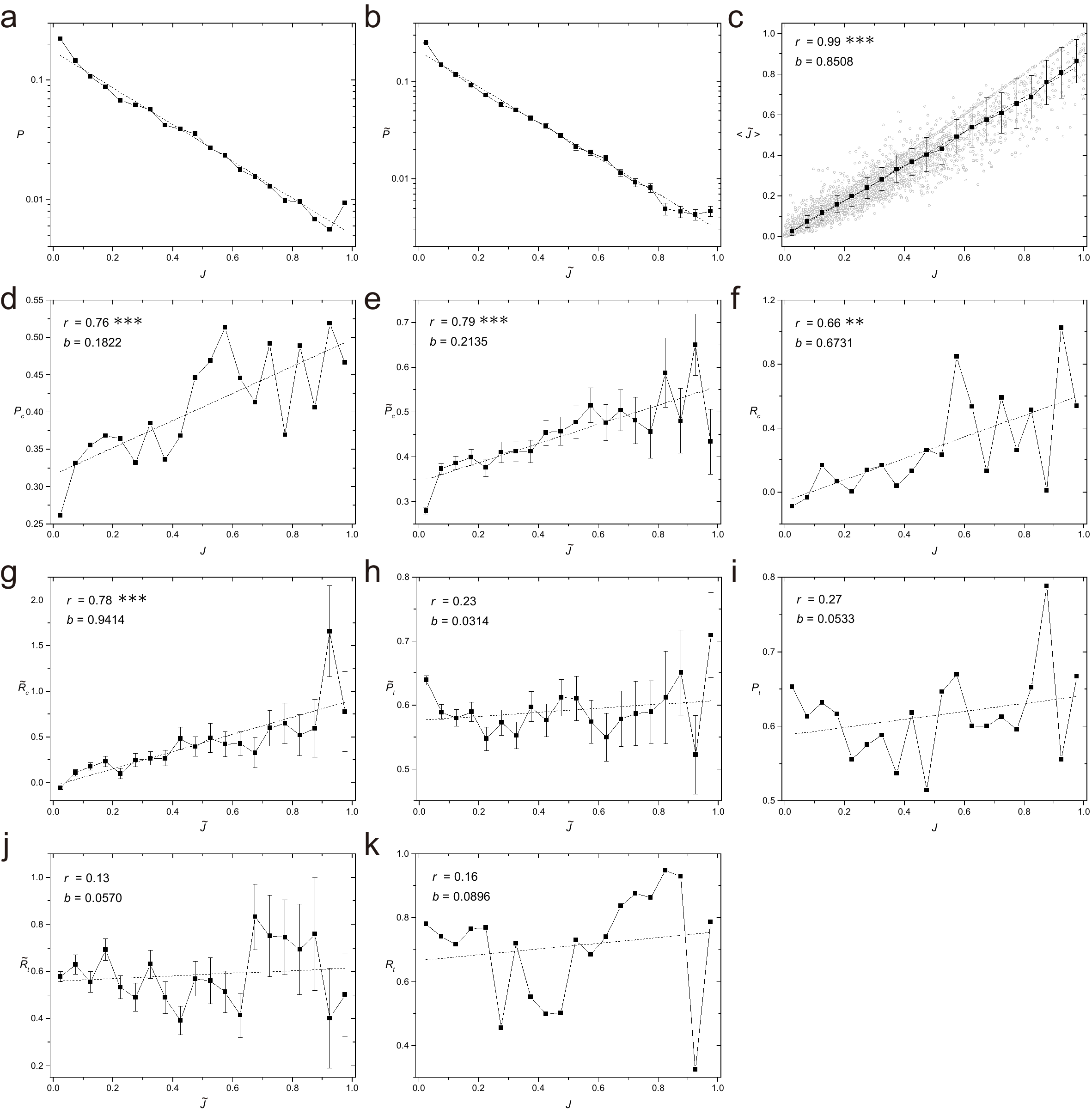}}
		\caption*{Figure S13:
			The results conditioning on the scientists whose $n$ falls in the range (15, 20].
			({\bf a-b}) The fraction of scientists $P$ and $\tilde{P}$ drop exponentially with $J$ and $\tilde{J}$.
			({\bf c}) For a scientist with $n$ papers, we calculate the average value $\langle \tilde{J} \rangle$ of her $n-1$ $\tilde{J}$. Then we plot her $\langle \tilde{J} \rangle$ versus $J$ (grey circle), and the mean value of $\langle \tilde{J} \rangle$ conditioning on the range of ($J-0.025,J+0.025$] (scatter with line). The result shows that $J$ and $\tilde{J}$ are consistent at the individual level.
			({\bf d-e}) $P_c$ and $\tilde{P}_c$ are positively correlated with $J$ and $\tilde{J}$.
			({\bf f-g}) $R_c$ and $\tilde{R}_c$ are positively correlated with $J$ and $\tilde{J}$.
			({\bf h-i}) $\tilde{P}_t$ and $P_t$ are not correlated with $\tilde{J}$ and $J$.
			({\bf j-k}) $\tilde{R}_t$ and $R_t$ are not correlated with $\tilde{J}$ and $J$.
			The value of $b$ is defined as the slope of the corresponding linear regression function (The dashed line). *** $p < 0.001$, ** $p < 0.05$, * $p < 0.1$ ($t$-test for Pearson coefficient $r$).}
	\end{center}
\end{figure}\noindent
\begin{figure}[h]
	\begin{center}
		\hypertarget{figS14}{\includegraphics[width=13.5cm]{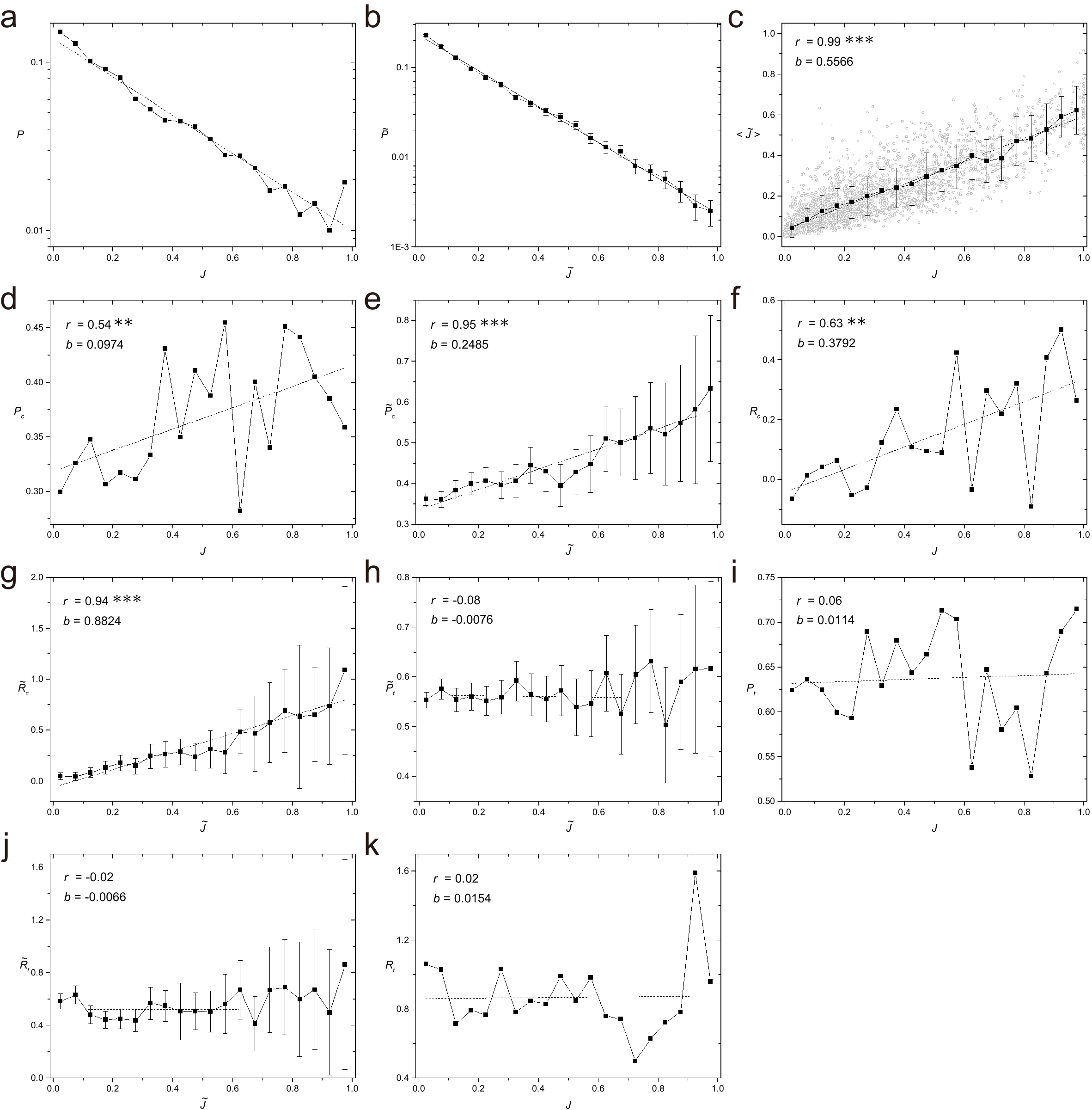}}
		\caption*{Figure S14:
			The results conditioning on the scientists whose $n$ falls in the range (20, 25].
			({\bf a-b}) The fraction of scientists $P$ and $\tilde{P}$ drop exponentially with $J$ and $\tilde{J}$.
			({\bf c}) For a scientist with $n$ papers, we calculate the average value $\langle \tilde{J} \rangle$ of her $n-1$ $\tilde{J}$. Then we plot her $\langle \tilde{J} \rangle$ versus $J$ (grey circle), and the mean value of $\langle \tilde{J} \rangle$ conditioning on the range of ($J-0.025,J+0.025$] (scatter with line). The result shows that $J$ and $\tilde{J}$ are consistent at the individual level.
			({\bf d-e}) $P_c$ and $\tilde{P}_c$ are positively correlated with $J$ and $\tilde{J}$.
			({\bf f-g}) $R_c$ and $\tilde{R}_c$ are positively correlated with $J$ and $\tilde{J}$.
			({\bf h-i}) $\tilde{P}_t$ for a range of values ($0 \leq \tilde{J} \leq 0.7$) with about 97\% of the sample size and small standard deviations, and $P_t$ are not correlated with $\tilde{J}$ and $J$.
			({\bf j-k}) $\tilde{R}_t$ for a range of values ($0 \leq \tilde{J} \leq 0.7$) with about 97\% of the sample size and small standard deviations, and $R_t$ are not correlated with $\tilde{J}$ and $J$.
			The value of $b$ is defined as the slope of the corresponding linear regression function (The dashed line). *** $p < 0.001$, ** $p < 0.05$, * $p < 0.1$ ($t$-test for Pearson coefficient $r$).}
	\end{center}
\end{figure}\noindent
\clearpage

\hypertarget{Note.S6}{\noindent{\bf S6. Correlation between scientific performance and $J$ caused by $n$}}
\\

We assume the linear dependency between $n$ and $J$, as well as  between $P_t$ and $n$. The relationship between $n$ and $J$ can be found via linear regression as $z = 28.96 + 10.943x$ (Fig. \ref{fig3}c). The relationship between $P_t$ and $n$ is $y = 0.5304 + 0.0048z$ (Fig. \ref{fig3}d). By combining both equations, we obtain the relationship between $P_t$ and $J$ as $y = 0.6694 + 0.0525x$, which has almost the same slop as that in the main text (Fig. \ref{fig3}b). It confirms that changing research interest alone does not hurt the growth fraction of productivity and the increasing trend is solely caused by the hidden variable $n$. Similarly, for results when $m = 7$, we have $y = 0.5088 + 0.0056z$ for relationship between $P_t$ and $n$ as well as $z = 26.183 + 10.385x$ for $n$ and $J$, giving rise to $y = 0.6554 + 0.0582x$. For results when $m = 9$, we have $y = 0.5353 + 0.0047z$ for relationship between $P_t$ and $n$ as well as $z = 31.638 + 11.95x$ for $n$ and $J$, giving rise to $y = 0.6840 + 0.0562x$.

When it comes to the linear dependency between $R_t$ and $n$ as well as $n$ and $J$. As well, the relationship between $n$ and $J$ is $z = 28.96 + 10.943x$ (Fig. \ref{fig3}c). The relationship between $R_t$ and $n$ is $y = 0.1952 + 0.0300z$ (Fig. \hyperlink{figS3}{S3}c). By combining both equations, we obtain the relationship between $R_t$ and $J$ as $y = 1.0641 + 0.3283x$, which is a little bigger than the slop in the Fig. \hyperlink{figS3}{S3}b). It confirms that changing research interest alone does not hurt the growth rate of productivity and the increasing trend is solely caused by the hidden variable $n$. Similarly, for results when $m = 7$, we have $y = 0.1774 + 0.0347z$ for relationship between $R_t$ and $n$ as well as $z = 26.183 + 10.385x$ for $n$ and $J$, giving rise to $y = 1.0860 + 0.3604x$. For results when $m = 9$, we have $y = 0.1376 + 0.0285z$ for relationship between $R_t$ and $n$ as well as $z = 31.638 + 11.95x$ for $n$ and $J$, giving rise to $y = 1.0393 + 0.3406x$.
\clearpage

\hypertarget{Note.S7}{\noindent{\bf S7. Measuring the diversity change}}
\\

We adopted two more types of diversity index to measure the diversity change of a scientist’s topic vectors (Table \hyperlink{tableS1}{S1}). As a higher Gini index indicates a higher level of inequality and a lower level of diversity, so we applied 1 – Gini in our analysis to represent the diversity. In addition, the Simpson index is widely used to measure the topic diversity, which captures both variety and balance of topics. The Simpson index formulated as $\sum_{j=1}^{n} x_j^2$ is negatively associated with diversity, where $x_j$ represents the fractionalized occurrence of $j^{th}$ PACS code in the topic vector $g$. Thus, we adopted $1-\sum_{j=1}^{n}x_j^2$ to characterize the topic diversity.
\begin{table*}[h]
	\centering
	\hypertarget{tableS1}{\caption*{Table S1: Diversity Measures}}
	\renewcommand\tabcolsep{1.7cm} 
	\begin{threeparttable}
		\begin{tabular}{cc}
			\hline  
			Measure & Description \\
			\hline \\ 
			1-Gini & $G=1-\frac{\sum_{j=1}^{n}(2j-n-1)\cdot x_j}{n\sum_{j=1}^{n}x_j} $ \\ \\
			Simpson index & $S=1-\sum_{j=1}^{n}x_j^2$ \\ \\
			\hline  
		\end{tabular}
		
		\begin{tablenotes}
			\footnotesize
			\item[*] The number $n$ of distinct PACS codes equals to 67, and $x_j$ represents the fraction of $j^{th}$ PACS code in the the topic vector $g$.  
		\end{tablenotes}
	\end{threeparttable}
\end{table*}
\newpage

We focus on scientists whose impact has increased and measure the diversity change of each individual. The results show that the scatters of diversity change are evenly distributed at both sides of $y=0$, and diversity change is uncorrelated with direction change (Fig. \hyperlink{figS15}{S15}). The fact that the change of diversity is close to 0 on average indicates that the increased impact is not from a more diverse or more narrow research agenda. Therefore, our finding is different from studies on interdisciplinary research.
\begin{figure}[h]
	\begin{center}
		\hypertarget{figS15}{\includegraphics[width=13.5cm]{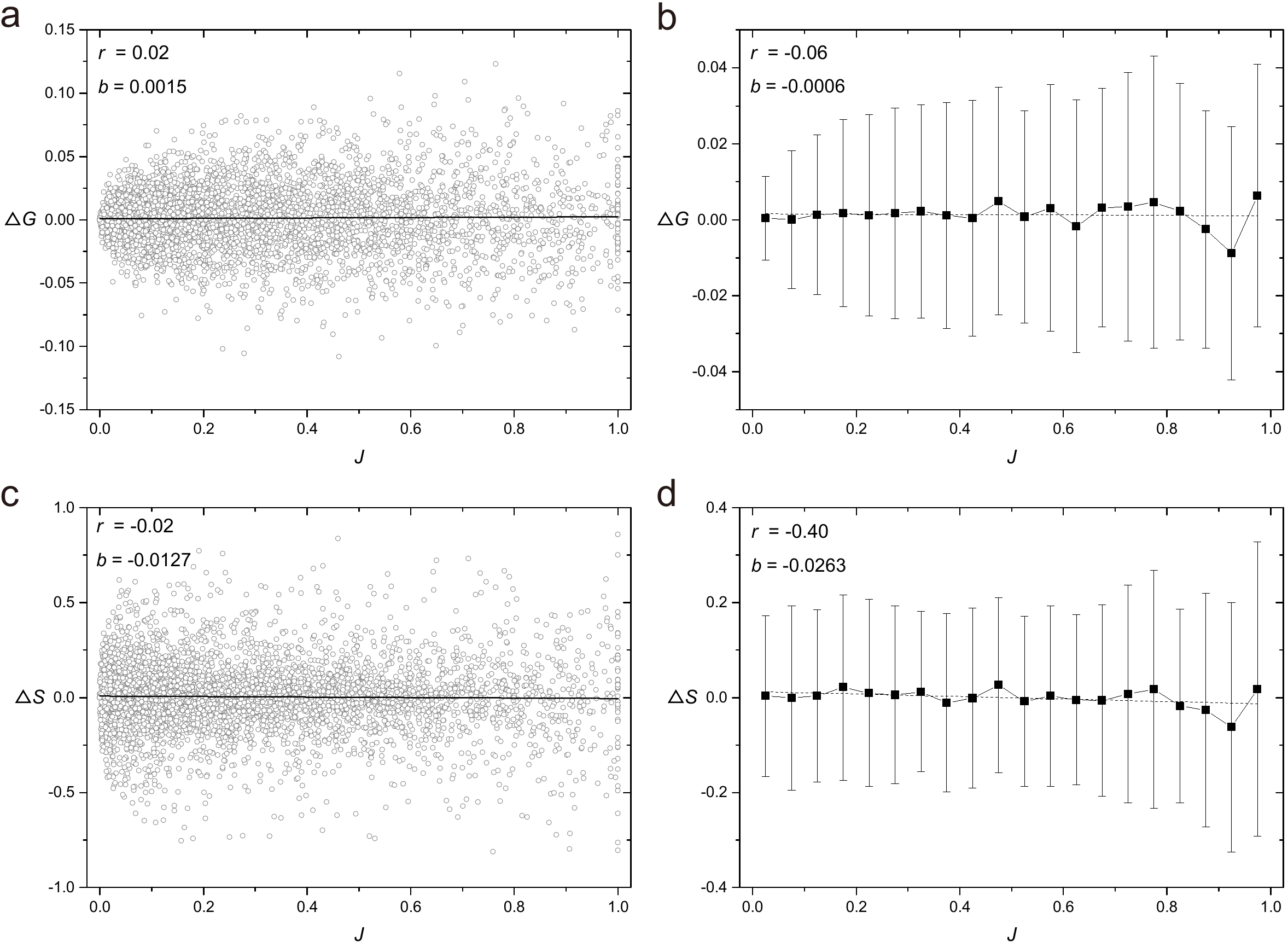}}
		\caption*{Figure S15:
			For scientists whose impact has increased ($\bar{c}_{2,f} > \bar{c}_{2,i}$), we plot their $J$ versus two types of diversity index, $\Delta G$ and $\Delta S$ respectively. 
			({\bf a}) The result shows that $\Delta G$ and $J$ are not correlated ($p>0.1$). The average $\Delta G$ is close to 0, indicating that diversity change is not associated with the impact increase.
			({\bf b}) Similar to ({\bf a}), but the mean value of $\Delta G$ is taken for individuals with ($J-0.025,J+0.025$].
			({\bf c}) As well, $\Delta S$ and $J$ are not correlated ($p>0.1$), and the average $\Delta S$ is close to 0.
			({\bf d}) Similar to ({\bf c}), but the mean value of $\Delta S$ is taken for individuals with ($J-0.025,J+0.025$].
			The value of $b$ is defined as the slope of the corresponding linear regression function. Error bars represent the one standard deviation of the mean.}
	\end{center}
\end{figure}\noindent

\end{document}